\documentclass[12pt,nofootinbib,notitlepage,preprint,aps,
  pra,
  amsmath,amssymb,longbibliography]{revtex4-2}
\usepackage{ulem}
\usepackage[varg]{txfonts}
\usepackage{bm}
\usepackage{float}
\usepackage{setspace}
\usepackage{xcolor}
\usepackage[left]{lineno}
\usepackage[caption=false]{subfig}
\PassOptionsToPackage{hyphens}{url}
\usepackage[bookmarks=true,         
     pdfnewwindow=true,      
     colorlinks=true,    
     linkcolor=blue,     
     citecolor=blue,     
     filecolor=blue,  
     urlcolor=blue,      
     final=true,
 ]{hyperref}
\usepackage{color,epsfig,amsmath,amssymb,fancyhdr} 

\usepackage{footnote}
\newcommand\footnoteref[1]{\protected@xdef\@thefnmark{\ref{#1}}\@footnotemark}
\makeatother
\usepackage{etoolbox}
\usepackage{footmisc}
\makeatletter
\patchcmd{\frontmatter@abstract@produce}
  {\vskip200\p@\@plus1fil
   \penalty-200\relax
   \vskip-200\p@\@plus-1fil}
  {}
  {}
  {}
\makeatother

\usepackage{booktabs}

\begin{document}
\singlespacing
\renewcommand{\abstractname}{}
\title{Layer formation in a stably stratified fluid cooled from above: Towards an analog for Jupiter and other gas giants}

\author{J. R. Fuentes$^1$}
\author{A. Cumming$^1$}
\author{E. H. Anders$^2$}

\affiliation{$^{1}$Department of Physics and McGill Space Institute, McGill University, Montreal, QC H3A 2T8, Canada\\
$^{2}$Center for Interdisciplinary Exploration and Research in Astrophysics, Northwestern University, Evanston, Illinois 60201, USA
}

\begin{abstract}
In 1D evolution models of gas giant planets, an outer convection zone advances into the interior as the surface cools, and multiple convective layers form beneath that convective front. To study layer formation below an outer convection zone in a similar scenario, we investigate the evolution of a stably-stratified fluid with a linear composition gradient that is constantly being cooled from above.  We use the Boussinesq approximation in a series of 2D simulations at low and high Prandtl numbers ($\mathrm{Pr} = 0.5$ and 7), initialized with constant temperature everywhere, and cooled at different rates. We find that multiple convective layers form at $\mathrm{Pr} = 7$, {as the result of an instability in the} diffusive thermal boundary layer below the outer convection zone. At low Pr, layers do not form because the temperature gradient within the boundary layer is much smaller than at large Pr and, consequently, is not large enough to overcome the stabilizing effect of the composition gradient. For the stratification used in this study, on the long-term the composition gradient is an ineffective barrier against the propagation of the outer convection zone and the entire fluid becomes fully-mixed, whether layers form or not. Our results challenge 1D evolutionary models of gas giant planets, which predict that layers are long-lived and that the outer convective envelope stops advancing inwards. We discuss what is needed for future work to build more realistic models.
\end{abstract}

\maketitle

\section{Introduction}

The Juno and Cassini missions have provided the best observational constraints on the internal structure of Jupiter and Saturn, both from measurements of their gravity fields and from seismology studies of Saturn’s rings \citep[see, e.g., ][]{2017SSRv..213....5B, 2017GeoRL..44.4649W,2020SSRv..216..122I, 2021NatAs...5.1103M}. These observations indicate that a large region within the interior of these planets (up to half of their radii) is likely stably-stratified by heavy elements, suggesting that their cores are diluted and more extended, without a sharp core-envelope transition as previously thought. 

This discovery has fundamental implications for the mixing processes at work in the interiors of gas giants. According to conventional models, gas giants have adiabatic interiors undergoing convection throughout almost the entire planet \citep{2016A&A...596A.114M}. However, composition gradients can significantly affect and even suppress convective motions \citep{1985Icar...62....4S,2012A&A...540A..20L}. Furthermore, under appropriate circumstances the interaction between the temperature and composition gradient leads to the formation of a staircase of turbulent convective layers separated by sharp interfaces across which transport of heat and chemical species is achieved by molecular diffusion  \citep{2018AnRFM..50..275G}.

A planet undergoing layered convection cools and mixes chemical elements less efficiently than a fully-convective planet, because transport is limited by diffusion between two adjacent convective layers. Layered convection has been proposed as a mechanism to explain several problems in planetary science. For example, it can reduce the rate of core erosion in Jupiter \citep{2017ApJ...849...24M}, explain the large radii of some extrasolar giant planets \citep{2007ApJ...661L..81C}, and explain the luminosity of Saturn, which is higher than predicted from fully-convective models \citep{2013NatGe...6..347L}. However, it is still not clear whether layered convection can persist over evolutionary time scales, which is essential for the validity of the solutions above.

Recent 1D evolutionary models of Jupiter with composition gradients find that a staircase of convective layers can form below the outer envelope, persisting over long timescales \citep{2018A&A...610L..14V,2020A&A...638A.121M,2022PSJ.....3...74S}. However, these layers could be an artificial effect due to the assumptions made in the models, which are restricted to 1D prescriptions for convective transport and convective boundary mixing. Because of this, the location and size of the layers depend on the number of grid points used in the simulations \citep{2018A&A...610L..14V}. Without resolving dynamics, it is unclear if the heavy elements are distributed within a extended stable region or over a staircase of multiple convective layers. Therefore, multi-dimensional numerical simulations are essential to understand the formation and evolution of convective staircases in compositionally-stratified fluids.

The improvement of computing capabilities has served as a bridge between hydrodynamic simulations and 1D evolution modeling. From the point of view of fluid dynamics, there are three important parameters that govern the dynamics of fluids with composition gradients: the Prandtl number $\mathrm{Pr}=\nu/\kappa_T$, which measures the ratio of kinematic viscosity $\nu$ to thermal diffusivity $\kappa_T$, the inverse Lewis number $\tau = \mathrm{Le}^{-1} = \kappa_S/\kappa_T$, which measures the ratio of the solute microscopic diffusivity $\kappa_S$ to the thermal diffusivity, and the ratio $\beta S_{z}/\alpha T_{z}$, which measures the stabilizing effect of the composition gradient with respect to the destabilizing effect of the temperature gradient (here, $\beta$ and $\alpha$ are the coefficients of solute contraction and thermal expansion, respectively). Under Jovian planet conditions, typical values for the microscopic diffusivities give $\mathrm{Pr} \sim 10^{-3} - 1$, and $\tau \sim 10^{-2}$ \citep{1977ApJS...35..221S,1977ApJS...35..239S,2012ApJS..202....5F}. Since we do not have a clear picture of Jupiter's interior, the exact value of $\beta S_{z}/\alpha T_{z}$ is unknown. However, it is expected to increase with depth from $\lesssim 1$ in the outer convection zone, towards $\gg 1$ in the core \citep{2017ApJ...849...24M}. 

Fortunately, the values of $\mathrm{Pr}$ and $\tau$ expected for Jupiter and other gas giants are accessible in numerical simulations. Recent 3D hydrodynamical simulations have shown that multiple convective layers can spontaneously form due to double-diffusive instabilities resulting from pre-existing temperature and composition gradients \citep{2011ApJ...731...66R,2012ApJ...750...61M,2013ApJ...768..157W,2016ApJ...823...33M}. These simulations have guided new transport prescriptions for 1D evolution models, but also have challenged them. For example, unlike the 1D models, 3D simulations have shown that convective staircases do not survive for a long time as the convective layers have a tendency to merge until a single fully convective layer remains.

Another complication is that it is not clear whether a staircase forms when the large-scale gradients develop over time. So far, numerical experiments that show layer formation are designed with idealized gradients to trigger the instabilities responsible for the formation of multiple layers. In planetary systems, the situation is different. The convective dynamics in gas giants are characterized by an outer convective envelope that advances into the core as the planet cools down, thus both the temperature and composition gradient evolve in time. Further, if a convective staircase forms below the outer envelope, it could be mixed or disrupted by overshooting motions beyond the bottom of the outer convection zone \citep[e.g.,][]{1997A&A...324L..81H,2022ApJ...926..169A}. This situation resembles experiments of water with a stable salinity gradient heated from below by applying a constant heat flux at the bottom \citep[e.g., ][]{1964PNAS...52...49T,turner_1968,fernando_1987}. In those experiments, multiple convective layers form successively from the bottom to the top of the fluid. However, those fluids are characterized by $\mathrm{Pr} = 7$, $\tau = 0.01$, and the dynamics observed there cannot be extrapolated to Jupiter's conditions. In this paper, we extend the laboratory experiments to fluids at lower Pr using 2D numerical simulations.

We investigate the long-term evolution of a fluid with a stable composition gradient that is constantly cooled from above. This mimics the evolution of gas giants, where a stable composition gradient (the core) opposes the inwards propagation of an outer convection zone (the envelope). Our goal is to see whether secondary convective layers can form and survive  under the vigorous mixing and turbulence of an outer convection zone at low Pr. 

This paper is organised as follows. Sect.~\ref{sec:model} describes the model and the numerical experiments conducted in this work.  In Sect.~\ref{sec:time_evol}, we provide a description of the dynamical evolution of the primordial composition gradient as a result of the inward propagation of the outer convection zone. We present analysis and results with emphasis on the differences between simulations in the parameter regime of astrophysical and geophysical flows (low and high Pr, respectively). In Sect.~\ref{sec:mecha_layers}, we analyse the structure of the thermal boundary layer underneath the outer convection zone, focusing on how Pr affects the thickness of the boundary layer, and the temperature step at the convective boundary. Both quantities are key to understanding the temperature gradient that is responsible for the formation of a second convective layer. We conclude in Section~\ref{sec:conclusions} with a summary and a general discussion.

\section{Details of the model and numerical method} \label{sec:model}

Since the convective dynamics described above requires very long integrations, we follow our previous work \citep{2020PhRvF...5l4501F} and perform two-dimensional simulations in a horizontally-periodic domain of height $H$ and width $L$, under the Boussinesq approximation \citep{1960ApJ...131..442S}. The density perturbations are small with respect to the background density of the fluid ($\rho/\rho_0 \ll 1$), and depend on the temperature and solute perturbations ($T$ and $S$, respectively) through $\rho = \rho_0 (\beta S - \alpha T)$ only in the gravity buoyancy term, where $\beta$ and $\alpha$ are the coefficients of compositional contraction and thermal expansion, respectively. We use $L=2H$ (aspect ratio $L/H = 2$) which is large enough to avoid the onset of artificial zonal flows in the fluid, which are known to suppress the vertical transport \citep{2014PhFl...26e4104F,2020JFM...905A..21W,2021PhRvF...6g4502F}. Convection is driven by a constant heat flux at the top boundary that cools down the fluid in time. Further, we use impermeable and stress-free top and bottom boundaries with no composition flux through them and no heat flux at the bottom.

We non-dimensionalize the Boussinesq equations using scales $H$, $H^2/\kappa_T$, $\kappa_T/H$, $S_{\mathrm{scale}}$,  $T_{\mathrm{scale}}$, for length, time, velocity, solute, and temperature, respectively. Here, $H$ is the height of the domain, and $\kappa_T$ is the thermal diffusivity. We set $S_{\mathrm{scale}}$ to the initial solute contrast across the box, and adopt $T_{\mathrm{scale}} = (\beta/\alpha) S_{\mathrm{scale}}$. Further, by this choice a unit of pressure corresponds to $\rho_0(\kappa_T/H)^2$.
The dimensionless equations are 
\begin{align}
&{\nabla} \cdot {\mathbf{v}} = 0\, ,\\ \label{eq:cont}
&\dfrac{\partial {\mathbf{v}}}{\partial {t}} +  ({\mathbf{v}}\cdot {\nabla}) {\mathbf{v}}  = - {\nabla} {P} + \mathrm{Pr}\mathcal{R}\left({T} - {S}\right)\hat{\mathbf{z}} + \mathrm{Pr} {\nabla}^2 {{\mathbf{v}}}\, ,\\ 
&\dfrac{\partial {S}}{\partial {t}} +  ({\mathbf{v}}\cdot {\nabla}){S}  = \tau {\nabla}^2 {S}\, ,\\
&\dfrac{\partial {T}}{\partial {t}} +  ({\mathbf{v}}\cdot {\nabla}){T}  =  {\nabla}^2 {T}\, . \label{eq:T}
\end{align}
There are 4 dimensionless numbers that govern the evolution of the flow. These are the Rayleigh, Prandtl, and inverse Lewis numbers, and the flux ratio (see discussion below). The Rayleigh, Prandtl, and inverse Lewis numbers are defined respectively as
\begin{align}
\mathcal{R} = \dfrac{ g\beta H^3 S_{\mathrm{scale}}}{\kappa_T \nu} \, ,\hspace{0.5cm} \mathrm{Pr} = \dfrac{\nu}{\kappa_T}\, ,\hspace{0.5cm} \tau = \dfrac{\kappa_S}{\kappa_T}\, .
\end{align}
Here $\kappa_S$ is the diffusivity of chemical species, and $\nu$ is the kinematic viscosity.  The parameter $\mathcal{R}$ is similar to the Rayleigh number in traditional thermal convection, but with an important difference that it measures the \textit{stability} of the fluid against convection. When fixing the size of the box, the gravity, and the thermodynamic properties of the fluid, an increase in $\mathcal{R}$ can be only due to an increase in the initial solute contrast, which results in an increase in the stability of the fluid against convection. Since we want to simulate a convection zone advancing into a stable region, a sufficiently large value of $\mathcal{R}$ can avoid overturning convection in the whole box.

The boundary conditions are 
\begin{gather}
 w\,\big\vert_{ z=0,1} = 0\,, \hspace{0.2cm}   \dfrac{\partial  u}{\partial  z}\,\,\bigg\vert_{ z=0,1} = 0\,, \hspace{0.2cm} \dfrac{\partial  S}{\partial  z}\,\bigg\vert_{z=0,1} = 0\,, \\
\dfrac{\partial  T}{\partial  z}\,\bigg\vert_{ z=0}  = 0\, , \hspace{0.2cm} \dfrac{\partial  T}{\partial  z}\,\bigg\vert_{ z=1}  = -\frac{F_0}{F_{\mathrm{crit}}}\, ,
\end{gather} 
where $w$ and $u$ are the vertical and horizontal velocity, respectively, and
\begin{equation}
F_{\mathrm{crit}} = k \frac{\beta}{\alpha}\frac{S_{\mathrm{scale}}}{H}\, ,
\end{equation}
is the critical heat flux \textit{across the box} that would make the fluid marginally stable against thermal convection. Here $k$ is the thermal conductivity.
The ratio $F_0/F_{\mathrm{crit}}$ is the fourth dimensionless parameter, and controls the rate at which the fluid cools down over time, given the initial composition gradient. 
Note that even values $F_0/F_{\mathrm{crit}} < 1$ can drive convection from the top (so $F_{\mathrm{crit}}$ should not be understood as a critical flux for the onset of convection in our time-dependent situation). This is because the temperature step between the top boundary and the fluid below increases as the fluid cools. However, the time that it takes for the fluid to be cold enough to overturn becomes larger as $F_0/F_{\mathrm{crit}}$ gets smaller. 

All the simulations in this work are initialized with the same stratification:   constant temperature $T_0 = 1$ everywhere, and a fixed solute profile that varies linearly with depth as $S_{0} = 1 - z$. This choice for the initial solute profile is inconsistent with the zero-flux boundary conditions for solute. However, this does not have a significant effect on our calculations because the running time of the simulations is much less than the time it takes for solute to diffuse across the box. Further, although the initial gradient is eroded near the top and bottom boundaries, it only slightly affects the bottom boundary. The convective motions near the top rapidly mix the initial gradient, making the solute concentration uniform everywhere inside the convection zone ($\partial S/\partial z = 0$, including the top boundary). Therefore, only the bottom boundary of the box is affected by producing a small erosion of the solute gradient, which does not affect our conclusions.

\begin{table}
\centering
\caption{Parameters used in the simulations. The second and third columns correspond to the Prandtl number and the flux ratio $F_0/F_{\rm crit}$ that sets the cooling boundary condition, respectively. The next column contains the simulation time in units of the thermal diffusion time across the box. All the experiments were conducted using a linear distribution of solute $S_0(z) = 1 - z$, constant temperature $T_0 = 1$, fixed diffusivity ratio $\tau=0.07$, and fixed $\mathcal{R}= 10^{10}$.} \label{tab_params}
\begin{tabular}{@{}lccccc@{}}
\hline
$\#$  & $\mathrm{Pr}$ & $F_0/F_{\rm crit}$ & $t_{\rm sim}$ [$t_{\rm diff}$]\\
\hline
1  & 0.5 & 0.5  & 0.720 \\ 
2  & 0.5 & 1   & 0.366 \\ 
3  & 0.5 & 5   &  0.075\\ 
4  & 0.5 & 10 & 0.028\\ 
5  & 7 & 1    & 0.885 \\ 
6 & 7 & 5   & 0.162 \\ 
7  & 7 & 10   & 0.058 \\ 
\hline
\end{tabular}
\end{table}
 
We solve Eqs. \eqref{eq:cont} -- \eqref{eq:T} using the spectral code Dedalus \citep{2020PhRvR...2b3068B}. The variables are represented on a Chebyshev (vertical) and Fourier (horizontally-periodic) domain. To avoid aliasing errors, we adopt the ``3/2 rule'' in both directions. This means the non-linear terms are evaluated on a spatial grid whose total number of points is $(3/2)^2$ greater than the number of mode coefficients ($N_x \times N_z$).  We use $N_x = 2048$ and $N_z = 1024$ modes in the horizontal and vertical directions, respectively. For timestepping, we use a third-order, four-stage, implicit-explicit Runge-Kutta scheme (RK443) \citep{ASCHER1997151}, where the linear and nonlinear terms are treated implicitly and explicitly, respectively. To start our simulations, we add random noise perturbations to the background temperature at the top boundary. 

Table \ref{tab_params} provides a list with the parameters used in the simulations. We use similar dimensionless parameters to the ones used in our previous work based on laboratory experiments \citep{2020PhRvF...5l4501F,2021PhRvF...6g4502F}. The simulations in this study are performed at fixed diffusivity ratio $\tau=\kappa_S/\kappa_T = 0.07$, and fixed Rayleigh number $\mathcal{R} = 10^{10}$. We set $\mathrm{Pr} = 0.5$, and vary the magnitude of the imposed cooling flux such that $F_0/F_{\rm crit} = 0.5$, 1, 5, and 10. We also compare our runs with a few selected simulations at $\mathrm{Pr}=7$. Finally, with the aim of studying the long-term behaviour of the fluid, we evolve the system until the whole box becomes fully-mixed.

\section{Dynamical evolution of the primordial composition gradient} \label{sec:time_evol}

In this section, we discuss the time evolution of the composition profile as the outer convection zone propagates inwards and eventually mixes the entire fluid. We first discuss the speed at which the convection zone moves inwards (Sect.~\ref{sec:prop_out}), the presence and lack of secondary layers at high and low Pr, respectively (Sect.~\ref{sec:2nd_layers}), and the profiles of composition fluxes (Sect.~\ref{sec:flux_profiles}).
\subsection{Propagation of the outer convection zone. Does the fluid become fully-mixed?} \label{sec:prop_out}

\begin{figure*}
\centering
\includegraphics[width=0.7\textwidth]{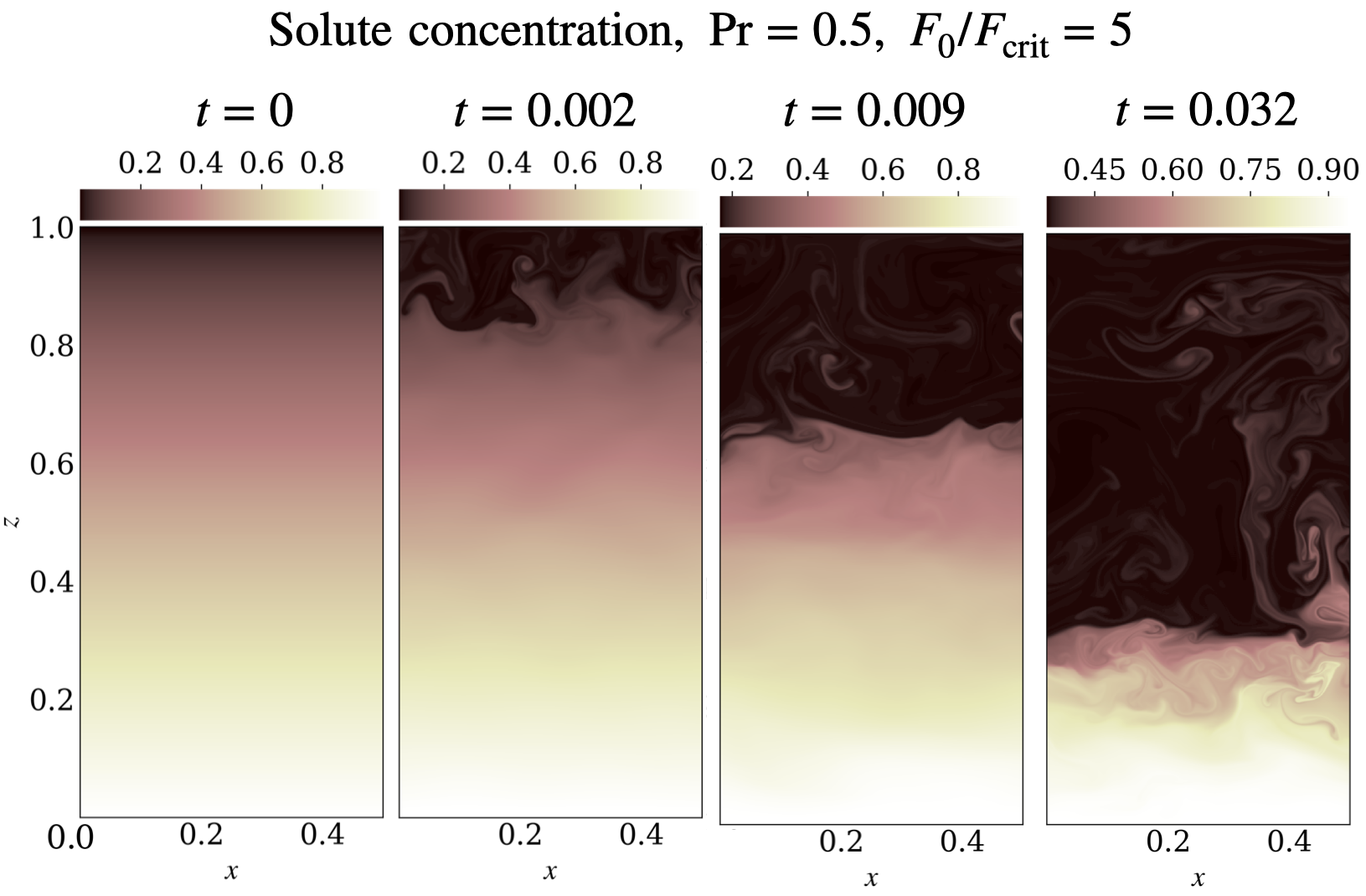}\\
\includegraphics[width=\textwidth]{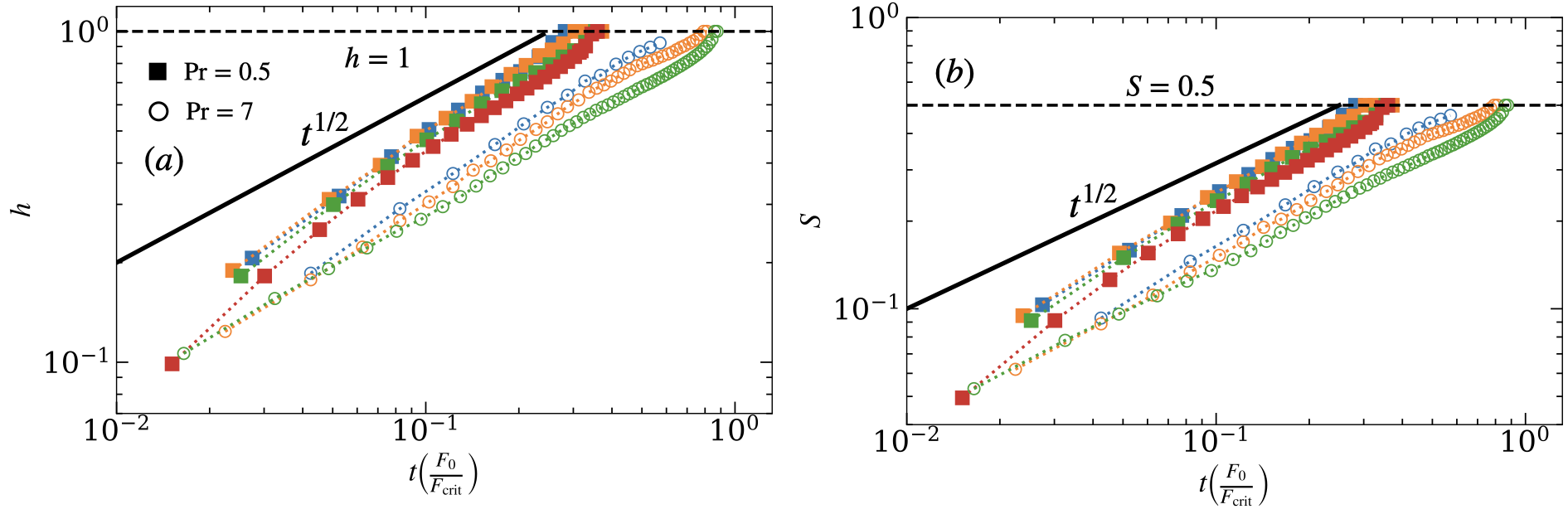}\\
\caption{Upper panel: 2D snapshots of the solute concentration at different times. Results are shown for the run using $\mathrm{Pr}=0.5$ and $F_0/F_{\rm crit} = 5$. Note that we only show the region delimited by $x\in 0-0.5$. Lower panel (a): Thickness of the outer convection zone, $h$, as a function of $t\, (F_0/F_{\rm{crit}})$, where $t$ denotes time. At each time, we measure the size of the outer convection zone as the distance between the top boundary and the location where the solute concentration varies at most by $5\%$ with respect to its value at the top boundary. Lower panel (b): Average solute concentration in the outer convection zone, $S$, as a function of $t\, (F_0/F_{\rm{crit}})$. As it is expected for a initial solute distribution of the form $S_0 = 1 - z$, when the entire box is fully-mixed the solute concentration is 0.5. In both panels, the results are shown for all simulations at $\mathrm{Pr}=0.5$ and $\mathrm{Pr}=7$, starting with a uniform temperature distribution. Data points correspond to direct measurements from the simulations. The solid line is the scaling $h\propto t^{1/2}$.} \label{fig_h_time}
\end{figure*}

After turning on the cooling flux at the top boundary, a thermal boundary layer develops and becomes unstable to convection. As the fluid cools over time, convective motions mix the primordial linear distribution of solute, forming a well-mixed convective zone layer on top of a stable fluid (Fig.~\ref{fig_h_time} upper panel). Previous studies have shown that the growth of the outer convection zone is due to eddies that overshoot into the stable region and entrain material from it \citep[e.g.,][]{fernando_1987,1997JFM...331..199J}. In \citet{2020PhRvF...5l4501F}, we studied in detail the evolution of the outer convection zone, with emphasis on the mixing processes at the convective boundary. We made an analytic model that predicts that the size of the convection zone evolves with time according to
\begin{equation}
h = \left(2C\right)^{1/2} \left(\frac{F_0}{F_{\rm crit}}\right)^{1/2} t^{1/2}\, ,\label{eq_h_prediction}
\end{equation}
(given our non-dimensionalization). The proportionality constant is $C = 1-\varepsilon + 2 \gamma$,  where $\gamma$ is the fraction of the kinetic energy flux available to mix material across the interface at the base of the convection zone (entrainment efficiency), and $\varepsilon$ is the ratio between the interfacial heat flux and the cooling flux at the top of the convection zone \citep[for more details, we refer the reader to][]{2020PhRvF...5l4501F}. Both parameters were measured from the simulations, giving $\gamma \sim 1$ at low Pr, whereas $\gamma \sim 0.1$ for $\mathrm{Pr} = 7$, indicating a higher mixing efficiency in fluids of low Pr. The effect of heat coming from below was much more weakly-dependent on Pr, with $\varepsilon \sim 0.3$ on average for both low and high Pr. The simulations in this work show the same trend (Fig.~\ref{fig_h_time}a), and as in our previous work, we find that fitting the convection zone thickness as a function of time gives deviations from the $t^{1/2}$ scaling, with $h(t)\propto t^{0.57-0.61}$ instead. For example, Fig.~\ref{fig_h_time} shows that once the outer convection zone gets thicker, its growth rate slightly decreases, but then increases again once the convection zone reaches the bottom boundary. The latter is because of the contribution of molecular diffusion of solute. As expected, the convection zone grows faster at low Pr. The difference between the low and high Pr curves in Fig.~\ref{fig_h_time} is consistent with the values of $\gamma$ and $\varepsilon$ measured in \cite{2020PhRvF...5l4501F}.


Note that evolution of the solute concentration in the outer convection zone shows the same trend as $h$ (Fig.~\ref{fig_h_time}b). This is because at a given time $t$, the amount of solute that is transported upward into the outer layer is $\Delta \overline{S} (t) = 0.5 |dS_0/dz|\,h (t)$ \citep[for details, see][]{2020PhRvF...5l4501F}. Further, as it is expected for an initial distribution of solute of the form $S_0 = 1-z$, once the box is fully-mixed the solute concentration in the entire box is uniform and equal to 0.5.

\subsection{Vertical distribution of solute and secondary convective layers}\label{sec:2nd_layers}

\begin{figure*}[h!]
\centering
\includegraphics[width=\textwidth]{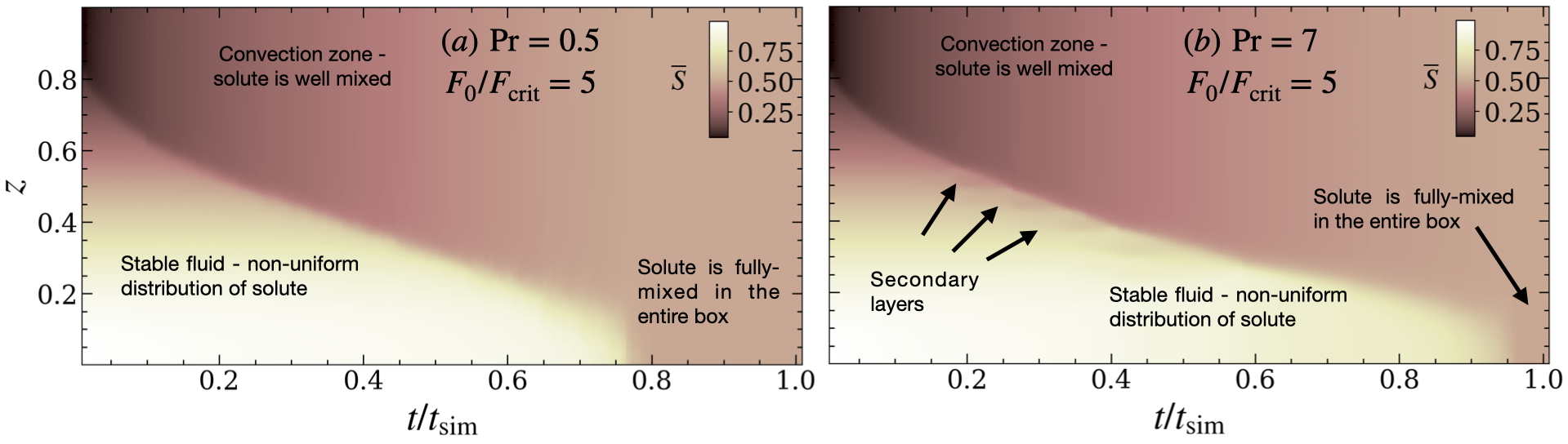}
\caption{Horizontally-averaged solute concentration at depth $z$ and time $t$ (normalized by $t_{\mathrm{sim}}$). Results are shown for two cases. Panels (a) and (b) show results for runs using $F_0/F_{\rm crit} = 5$  at $\mathrm{Pr}=0.5$ and $\mathrm{Pr}=7$, respectively. Note that at high Pr there are additional convective layers below the outer convection zone. These layers can persist in time but eventually they are engulfed by the outer convection zone.}
\label{fig:summary}
\end{figure*}

Although the final state of the fluid is the same for all the simulations, the evolution of the solute distribution depends on Pr. Fig.~\ref{fig:summary} summarizes the time evolution of the vertical profile of the solute concentration for selected simulations at low and high Pr. Initially, there is a linear distribution of solute across the box. As the convection zone propagates inwards, it mixes the solute and the concentration becomes uniform within the convective layer. At low Pr (panel a), the convection zone continuously mixes the solute gradient until the entire box is fully mixed. On the contrary, at high Pr there are time spans where the fluid develops secondary convective layers (panel b). However, the secondary layers do not stop the growth of the outer convection zone and the whole box becomes fully-mixed, as at low Pr.

\begin{figure*}
\centering
\includegraphics[width=0.8\textwidth]{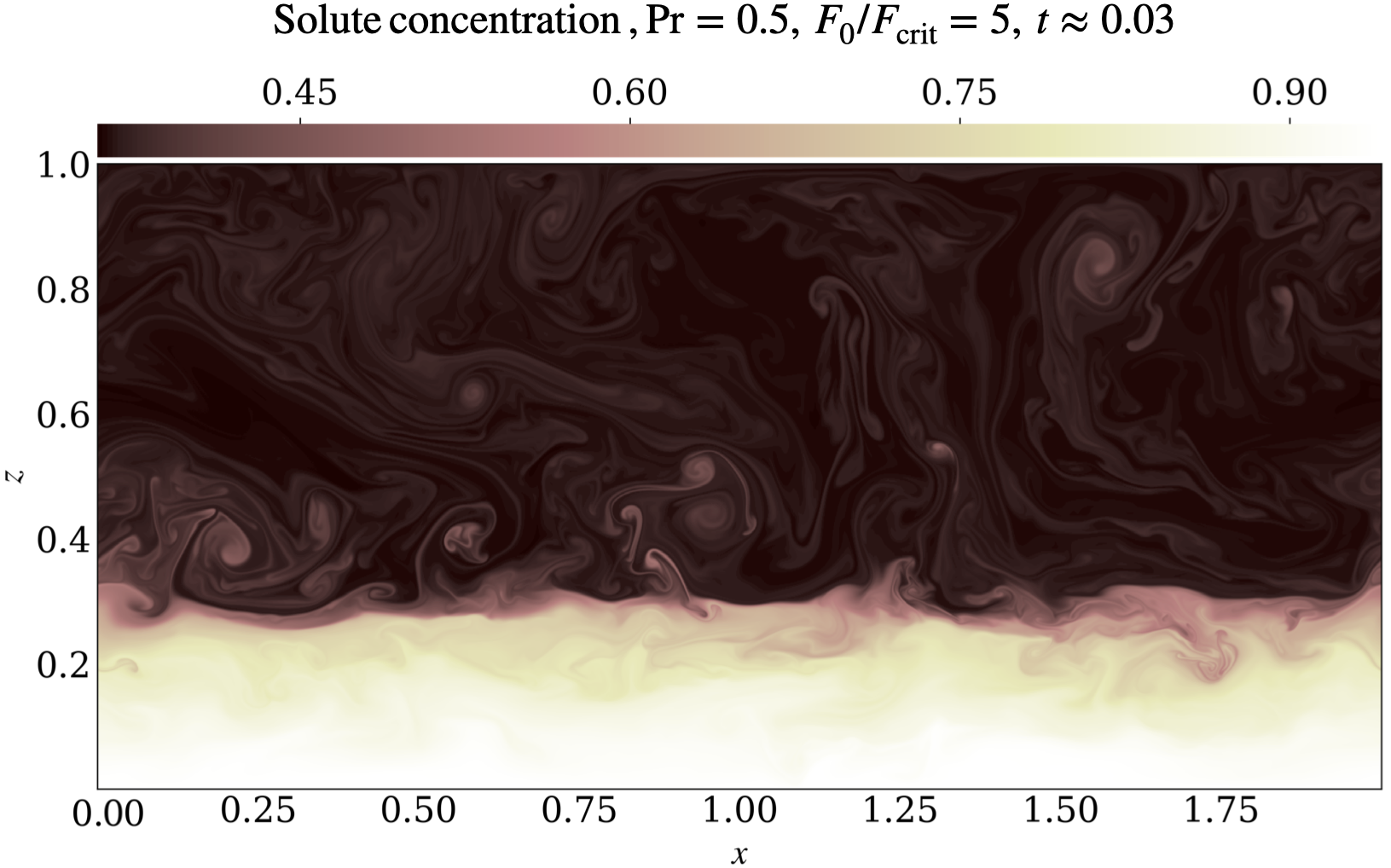}\\\vspace{0.5cm}
\includegraphics[width=0.8\textwidth]{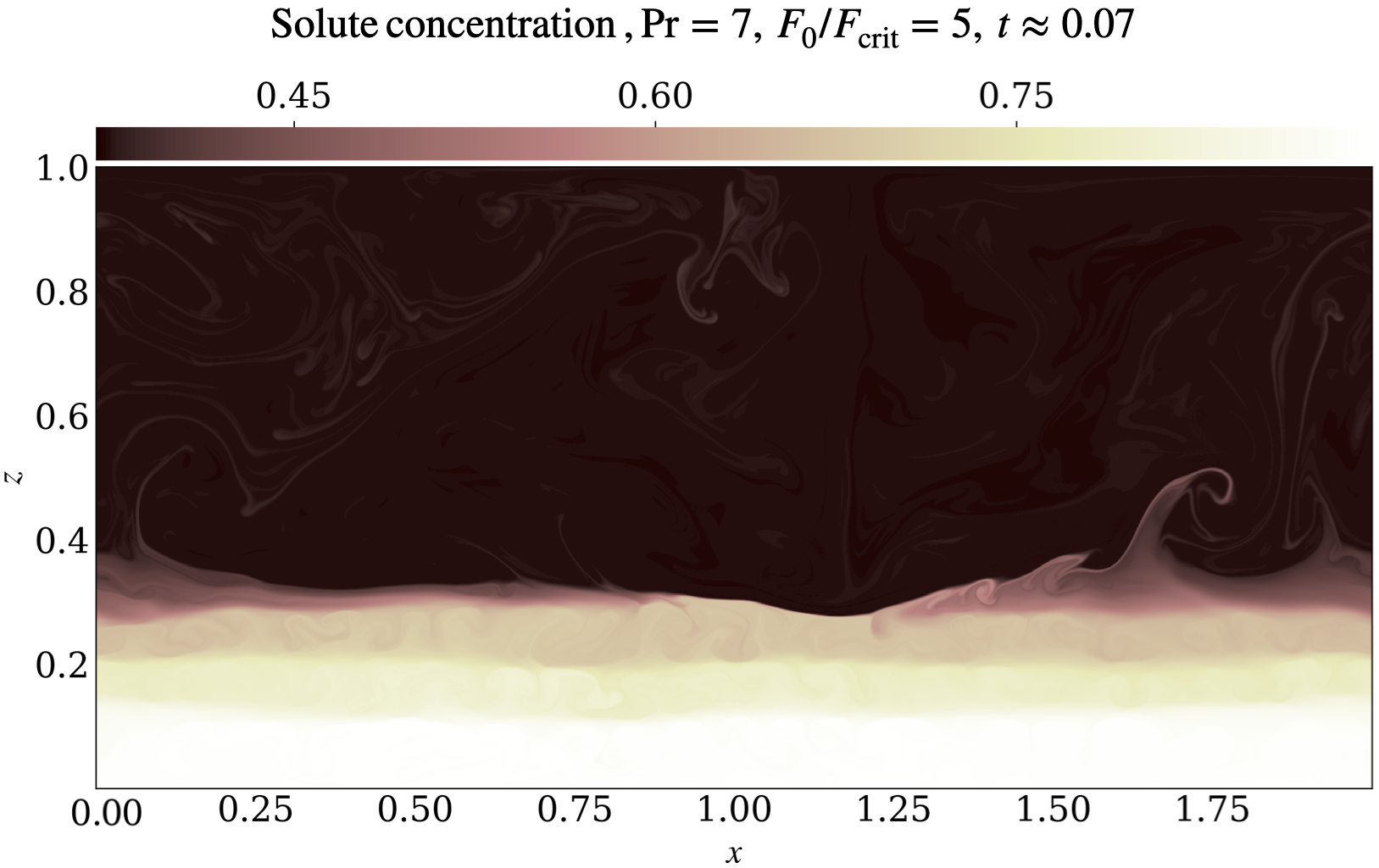}
\caption{2D snapshots of the solute field for runs using $\mathrm{Pr}=0.5$, $F_0/F_{\rm crit}=5$ (top panel) and $\mathrm{Pr}=7$, $F_0/F_{\rm crit} = 5$ (bottom panel). The fields are shown at times when the thickness of the convection zone is roughly the same for both simulations. The fluid at $\mathrm{Pr}=7$ exhibits secondary convective layers, whereas at $\mathrm{Pr}=0.5$ there are no clear secondary layers.}\label{fig_S_layers}
\end{figure*}

The secondary layers observed at high Pr can be seen more clearly in Fig.~\ref{fig_S_layers}, which shows 2D snapshots of the solute field for selected cases at low and high Pr. For better comparison, the snapshots were chosen at times where the thickness of the outer convection zone is roughly of the same size. At low $\mathrm{Pr}$, we observe the outer convection zone with a diffuse distribution of solute below. Although partially-mixed regions that resemble layers can be seen, e.g. between $x\sim$ $1.5$--$1.75$, they are short-lived. On the contrary, at high $\mathrm{Pr}$, below the outer convection zone, the solute is distributed over 3 additional convective layers that are stable over time and span the entire horizontal domain.

\begin{figure*}
\centering
\includegraphics[width=\textwidth]{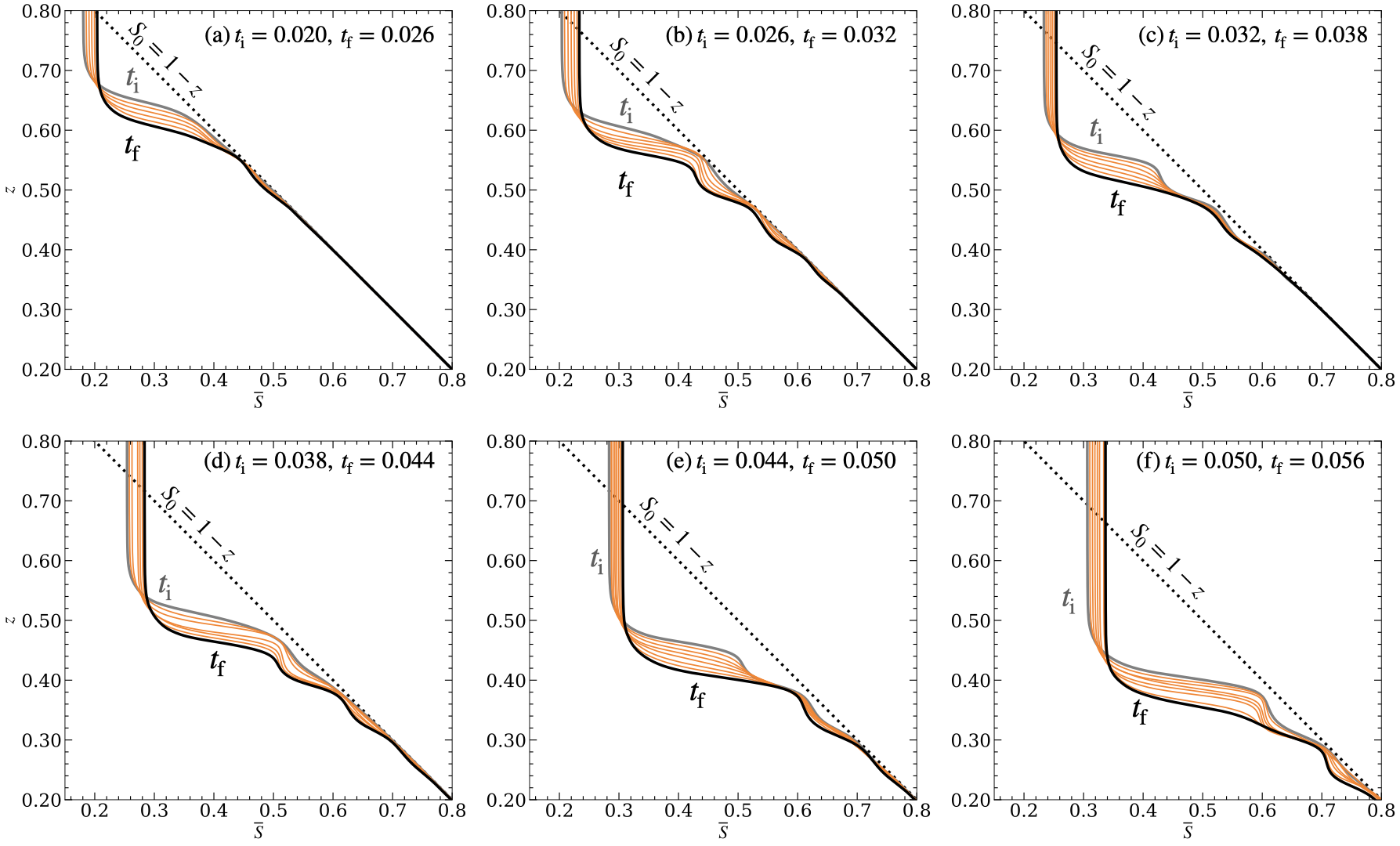}
\caption{Solute profiles for the run using $\mathrm{Pr}=7$ and $F_0/F_{\rm crit}=5$. Different panels show profiles at different times, as shown in the legends. The dotted line corresponds to the solute profile at $t=0$, whereas the gray and black solid lines correspond to profiles at $t_{\rm i}$ and $t_{\rm f}$, respectively. In all panels the outer convection zone propagates from top to bottom. Note that a convective staircase (secondary convective layers characterized by step-like structures) is visible in panels b-f. The solute concentration is roughly constant within the secondary convective layers (vertical regions), and undergoes a steep variation within the interfaces (horizontal regions). Due to the steep gradient, the interfaces are stable against convection and heat and solute are transported therein by diffusion.}
\label{fig:S_profiles}
\end{figure*}

The dynamics of the convective layers are clear when looking into vertical (horizontally-averaged) profiles of solute $\overline{S}(z)$ (the overbar on top means the quantity was averaged over the horizontal direction). Fig.~\ref{fig:S_profiles} shows $\overline{S}(z)$ at different times for the layering case above ($\mathrm{Pr}=7$ and $F_0/F_{\mathrm{crit}} = 5$). At early times (panel a), the concentration of solute in the outer convection zone increases as the front propagates inwards. The solute concentration transitions steeply between the bottom of the outer convection zone and the motionless fluid below. At the latest time in panel (a), we see a small amount of mixing at $z\approx 0.55$, below the solute step (interface). As time passes, we see a clear second convective layer where solute is well mixed, and a third one starts to develop at $z\approx 0.475$ (see latest profile in panel b). The convective flow in the secondary layer could in principle reduce the growth of the outer convective layer. However, we find the outer convection zone keeps moving inwards. Eventually the second layer is engulfed by the growing outer convection zone, but the third layer (now the second one below the outer convection zone) continues developing in its original position (as shown in panel c). With more evolution, the third layer is well mixed and a fourth layer at $z \approx 0.35$ starts to develop (panel d). As before, the third layer becomes engulfed by the outer convection zone while the fourth layer continues mixing the solute in its original position (panel e). The process repeats and when the fourth layer becomes well mixed, a fifth layer starts to form at $z\approx 0.27$. Eventually, the fourth layer becomes engulfed by the outer convection zone (panel f). This process repeats until the whole box becomes fully mixed.

\begin{figure*}
\centering
\includegraphics[width=\textwidth]{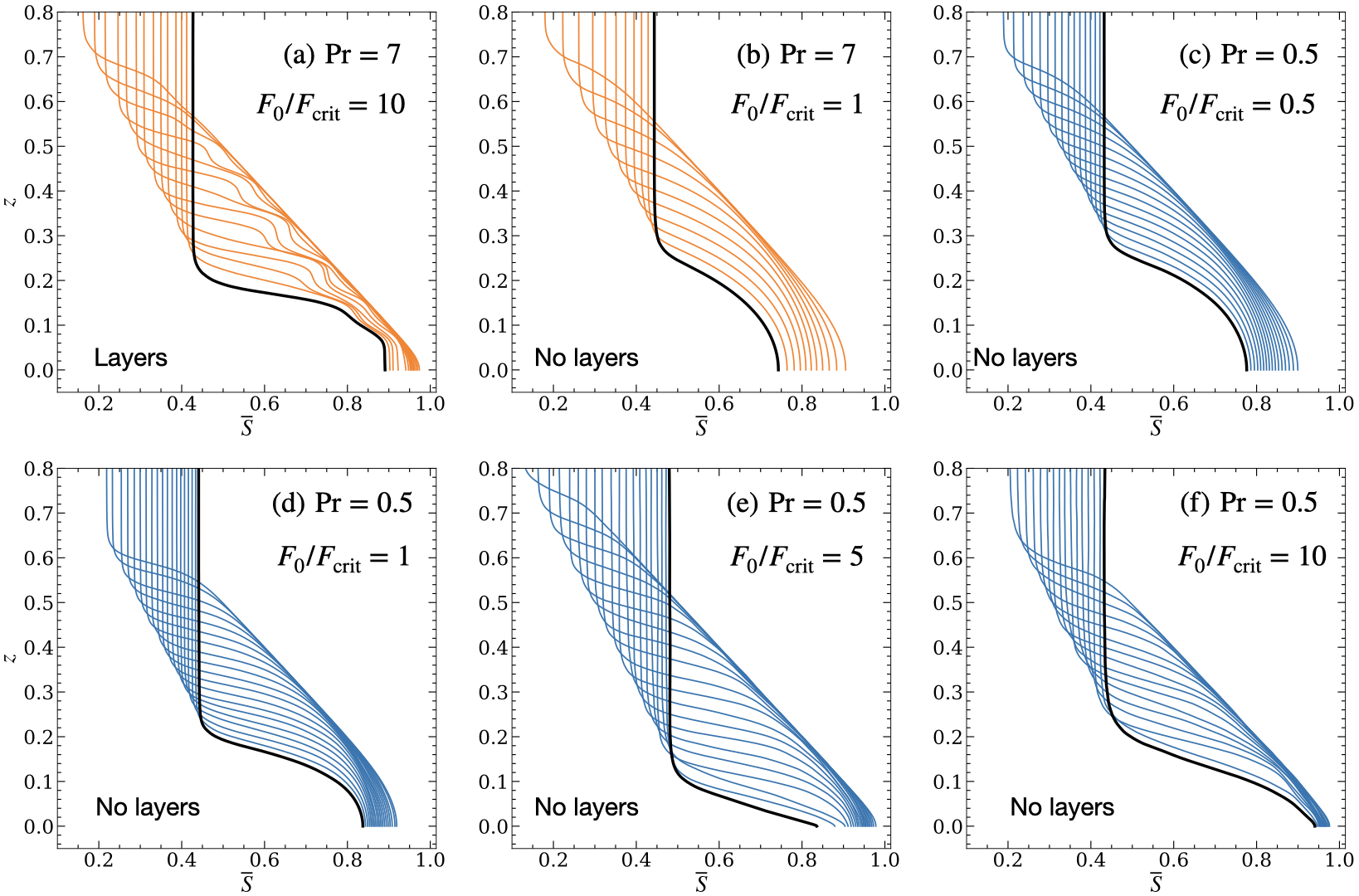}
\caption{Solute profiles for runs at $\mathrm{Pr}=7$ using $F_0/F_{\mathrm{crit}}=10$ and 1 (panels a and b, respectively). Panels (c)-(f) shows profiles for runs at $\mathrm{Pr}=0.5$ using $F_0/F_{\mathrm{crit}}=0.5$, 1, 5, and 10, as shown in the legends.  The black solid line corresponds to the latest profile among the ones shown. In all panels the outer convection zone propagates from top to bottom. Note that a convective staircase (secondary convective layers characterized by step-like structures) is visible at some times in panel (a). The solute concentration is roughly constant within the secondary convective layers (vertical lines), and undergoes a steep variation within the interfaces (horizontal lines).}
\label{fig:S_profiles_rest}
\end{figure*}

We find the same dynamics in the run using $\mathrm{Pr} = 7$ and $F_0/F_{\mathrm{crit}} = 10$, i.e., secondary convective layers develop and become engulfed by the outer convection zone (Fig.~\ref{fig:S_profiles_rest}a). However, when driving the system with a smaller cooling flux, e.g., the run using $F_0/F_{\mathrm{crit}} = 1$, we do not observe layer formation and the profiles are always composed of a well mixed region in the outer convection zone, and a steep transition to the primordial profile in the motionless fluid (Fig.~\ref{fig:S_profiles_rest}b).  Interestingly, we do not see layer formation in any of the runs at low Pr, no matter the magnitude of $F_0/F_{\mathrm{ crit}}$ (Fig.~\ref{fig:S_profiles_rest}c-f).  We explain these differences in Sect.~\ref{sec:mecha_layers}, where we analyse and discuss the mechanism for layer formation.

\begin{figure*}
\centering
\includegraphics[width=\textwidth]{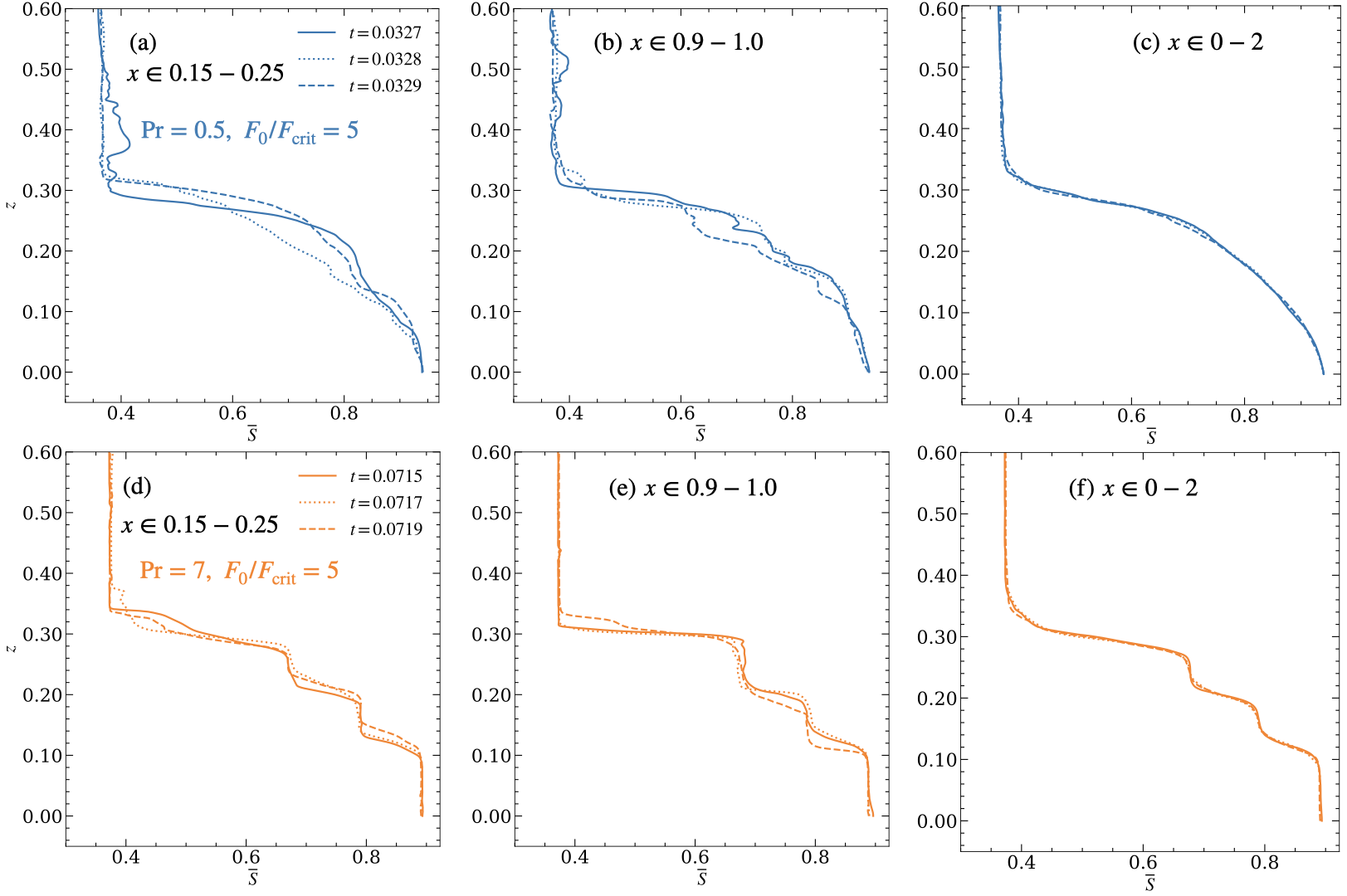}
\caption{Solute profiles for runs using $\mathrm{Pr}=0.5$ and $F_0/F_{\rm crit}=5$ (panels a-c)  and $\mathrm{Pr}=7$ and $F_0/F_{\rm crit}=5$ (panels d-f). For each panel, different curves distinguish between profiles taken at different times, separated by time intervals of $\Delta t \sim 10^{-4}$ (less than a percent of the simulation time, see Table \ref{tab_params} for the simulation time of each run). Panels in the first two columns show profiles computed from horizontally averaging over slices of width $\Delta x = 0.1$ in the horizontal direction. Panels in the right column show profiles computed from averaging over the whole horizontal extent of the box.}
\label{fig:S_averages}
\end{figure*}

We have investigated whether the horizontal averaging washes out the steps in the solute profiles of low Pr runs, thus explaining the lack of convective staircases at low Pr. We do so by averaging over small intervals across the horizontal direction when computing the vertical profiles.  Fig.~\ref{fig:S_averages}a-c shows step-like features at low Pr, but they depend strongly on the location where the horizontal average is taken, and they vary strongly over short time scales. We see that averaging over the entire $x$-domain smooths out the curves and the steps disappear (see panel d). This is not the case for the high Pr case, where the steps are visible regardless of where the horizontal average is taken and do not exhibit a significant variability over short timescales (panels e-h). There is therefore a clear difference between the low and high Pr cases; only the high Pr simulations show sustained global layers. We examine flux profiles in these regions in Sect.~\ref{sec:flux_profiles}.

\subsection{Solute fluxes} \label{sec:flux_profiles}

The differences between low and high Pr runs described above can also be observed in the profiles of solute flux\footnote{The shape and dynamics of the heat fluxes resembles that of the solute fluxes. Since in this section we focus on the evolution of the composition across the box, we just show the solute fluxes.}, which in dimensionless units is defined  as
\begin{align}
\overline{F}_S = \overline{wS} - \tau d\overline S/dz\, , \label{eq_solute_flux}
\end{align}
where the first and second term on the right hand side of Eq.~\eqref{eq_solute_flux} correspond to the advective and diffusive flux, respectively. Note that by the nondimensionalization described above, the solute flux is normalized to $(\kappa_T S_{\mathrm{scale}}/H)$.

\begin{figure*}
\centering
\includegraphics[width=0.9\textwidth]{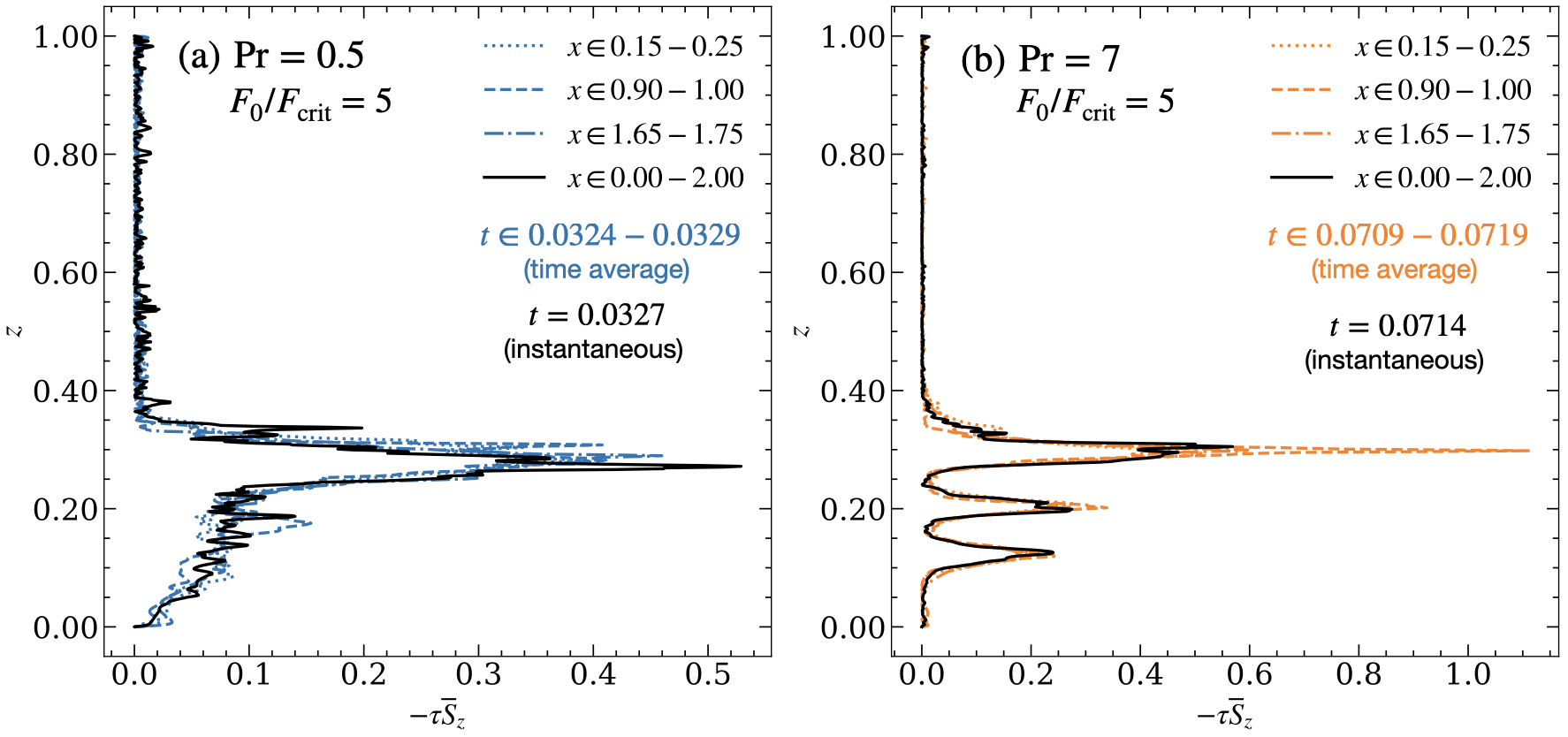}
\caption{Vertical profiles of the solute flux due to diffusion diffusion for the cases $\mathrm{Pr} = 0.5$ (panel a) and $\mathrm{Pr} = 7$ (panel b), with $F_0/F_{\mathrm{crit}} = 5$. In each panel, the color curves were made by averaging over small horizontal slices and short time intervals (as shown in the legends). The black curves correspond to profiles where the horizontal average consider the entire width of the box, and at an instantaneous time, as shown in the legends. The diffusive interfaces are the regions where the flux exhibits a Gaussian-like peak. Between interfaces, diffusion falls to zero, indicating a well-mixed region (convective layer). No signature of secondary layers or staircases are observed at low Pr. }
\label{fig:local_gradients}
\end{figure*}

As for the solute profiles, the horizontal average is also a concern for the diffusion fluxes. Since the interfaces are not flat, by performing horizontal averages the vertical gradients are smoothed out and the diffusive fluxes across interfaces could not be represented well. However, we find that horizontal averaging effects lower the measured fluxes by less than a factor of 2. We compute profiles of the diffusion flux of solute for different slices in the $x$ direction, and average them over short timescales (Fig.~\ref{fig:local_gradients}). The profiles are computed for the same time snapshot as Fig. \ref{fig:S_averages}. We observe the time-average (over a short time) of the diffusion flux from a slice does not vary significantly from the spatial average over the entire $x$ direction at a given time. Also, for the large Pr case, the diffusion flux profile has a well defined shape when a convective staircase is present (diffusion peaks at the separating interfaces, and decays to zero between the interfaces, indicated a well-mixed convective region). This is not the case for low Pr, since below the first diffusive interface the flux remains non-zero. Therefore, the steps observed locally at low Pr are not convective staircases.

\begin{figure*}
\centering
\includegraphics[width=16cm]{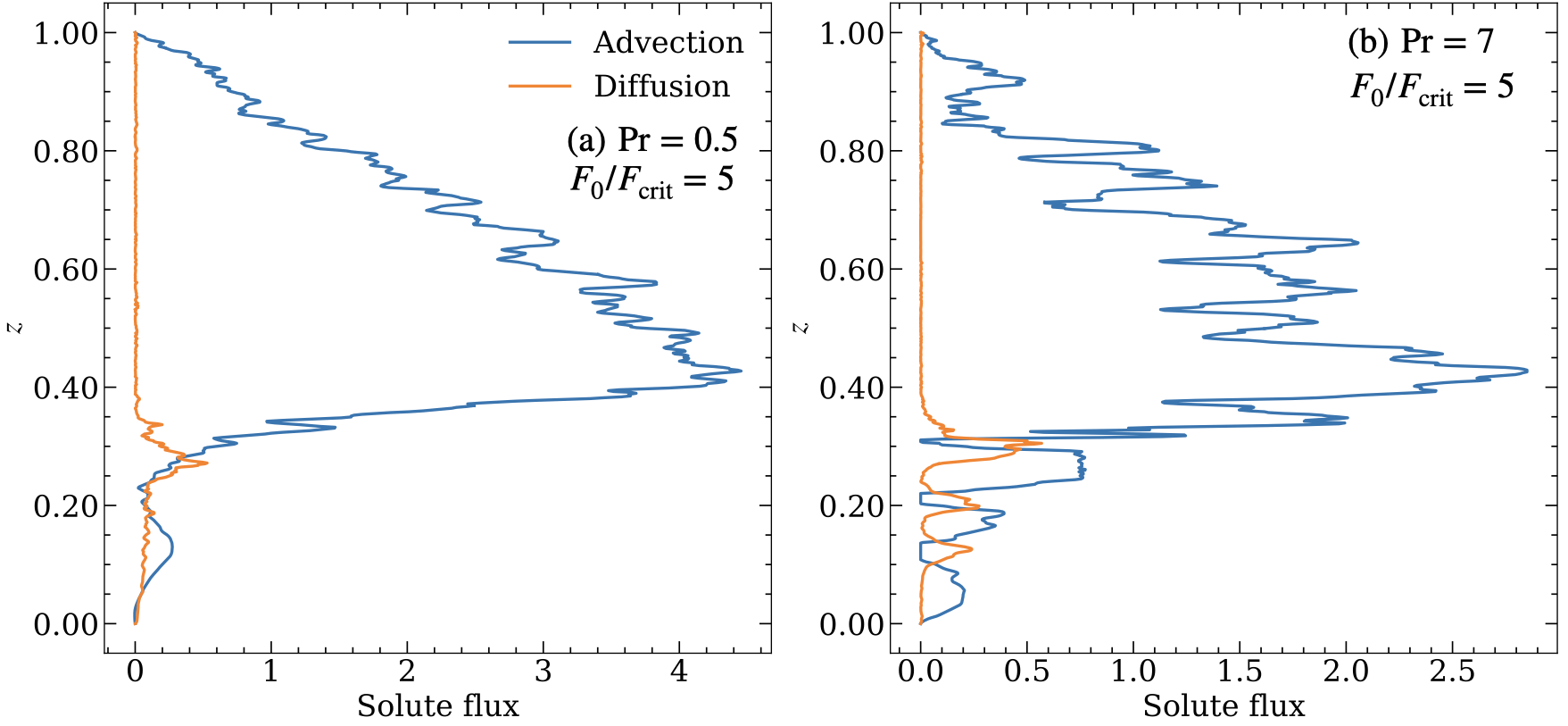}
\caption{Horizontally averaged flux profiles of solute at a time when the convection zone has advanced to $z\approx 0.4$. Panels (a) and (b) show profiles the runs using $\mathrm{Pr}=0.5$ and $\mathrm{Pr}=7$, using $F_0/F_{\rm crit}=5$. In both panels, the blue and orange curves correspond to the advective and diffusive contribution to the flux, respectively.}
\label{fig_flux_profiles}
\end{figure*}

From now on, when presenting and discussing vertical profiles, we mean averaged quantities over the entire horizontal extent of the box, at an instantaneous time snapshot. Fig.~\ref{fig_flux_profiles} shows profiles of the solute fluxes for simulations at low Pr and high Pr (panels a and b) at a time when the outer convection zone has advanced to $z\approx 0.4$. We select the runs using $F_0/F_{\mathrm{crit}}$ = 5 since for this value of the cooling flux, convective-staircases form at high Pr. For the cases at low Pr (panel a) the solute flux is dominated by advection in outer the convection zone, and by diffusion at the interfaces ($z\approx 0.3$). Note that within the outer convection zone ($z \approx 0.4 - 1$), the convective flux of solute decreases linearly with $z$, increasing the solute content everywhere (in the convection zone) at a constant rate to keep its composition uniform. The case at high Pr is different (panel b). In addition to the linear region corresponding to the flux across the outer convection zone, we observe that there a few locations ($z\approx 0.1$, 0.2, and 0.3) where the contribution from diffusion dominates over advection. These are diffusive interfaces. Between the interfaces, advection dominates in a well-mixed region. We observe similar profiles for runs using $F_0/F_{\mathrm{crit}}$ = 10. 

\section{Mechanism for layer formation} \label{sec:mecha_layers}

In the previous section, we showed that long-lived secondary layers form at high Pr, but not at low Pr. We now investigate the formation mechanism for these layers and the reason for the difference between low and high Pr.

\subsection{Formation of secondary layers by instability of the thermal boundary layer}

The formation of multiple convective layers has been investigated extensively in laboratory experiments of salt-stratiﬁed water heated from below \citep[e.g.,][]{1964PNAS...52...49T}.  In those experiments, a convective layer forms and grows upward from the bottom of the container. As time passes, heat diffuses quickly through the top of the layer, while salt diffuses much more slowly, preserving the stability of the interface. Eventually, the thermal boundary layer ahead of the bottom convection zone becomes unstable according to the Ledoux criterion for convection, and forms a second convective layer.

\begin{figure*}
\centering
\includegraphics[width=\textwidth]{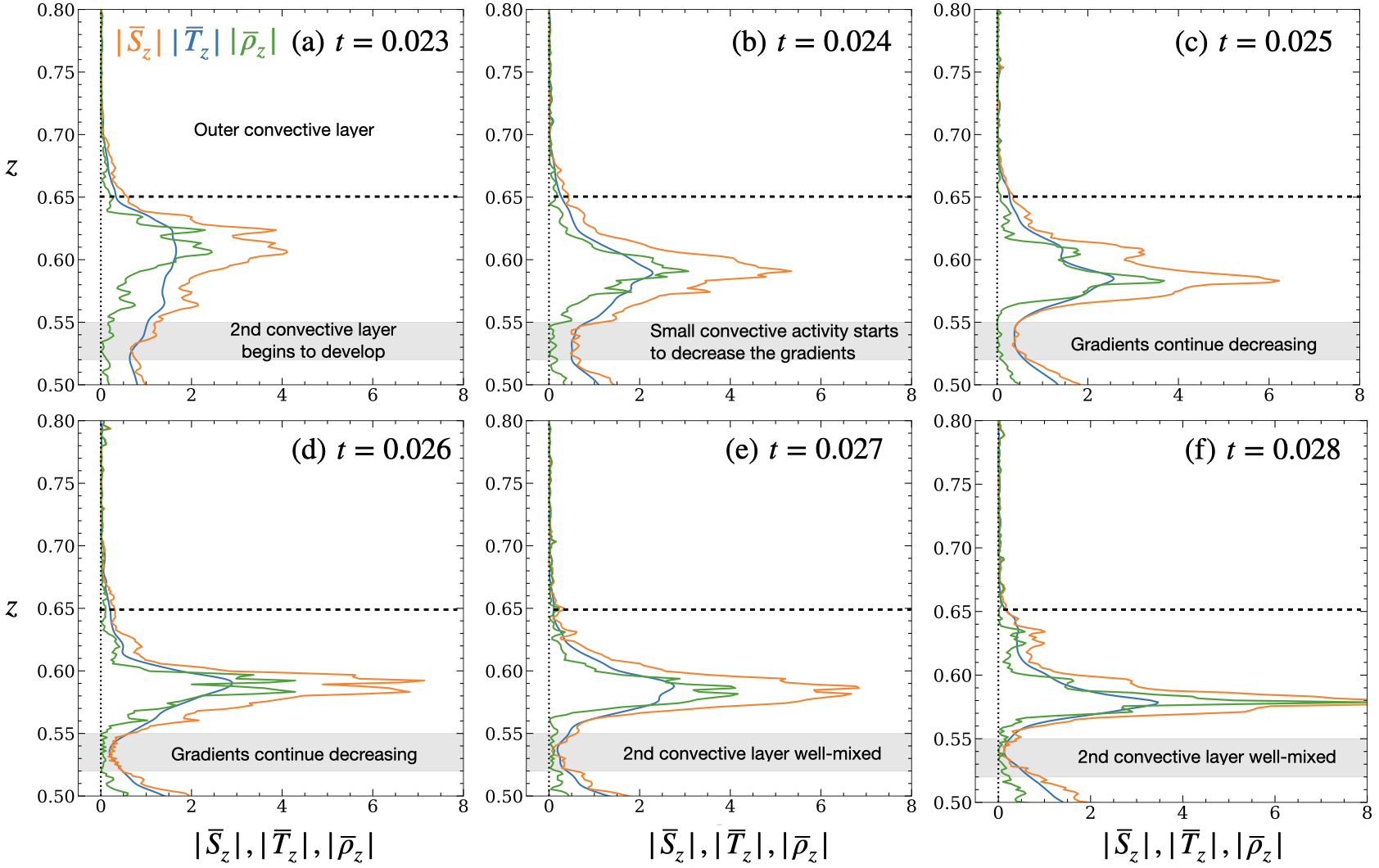}
\caption{Panels (a)-(f) show vertical profiles of $|\overline{S}_z|$, $|\overline{T}_z|$, and $|\overline{\rho}_z|$ (solute, temperature, and density gradients, respectively). Each panel shows the profiles at a particular time, to capture the behaviour of the gradients and solute during the formation of a second convective layer in the region defined by gray area. The results correspond to the run using $\mathrm{Pr} = 7$ and $F_0/F_{\rm crit} = 5$. The outer convection zone is the region between $z \approx 0.65$--$1$ (delimited by the dashed-line) where all the gradients are approximately zero.}
\label{fig:gradients}
\end{figure*}

Figure \ref{fig:gradients} shows profiles of the vertical gradients of solute, temperature and density, during the formation of a second layer in the run using $\mathrm{Pr} = 7$ and $F_0/F_{\rm crit} = 5$. When the magnitude of the temperature gradient becomes comparable to the magnitude of the solute gradient (i.e., when the Ledoux criterion for convective instability is satisfied), a second convective layer begins to develop in the region shaded in gray. Note that during the formation of the second layer, the density gradient is close to zero, even when the fluid is not well mixed (panels a-c). As convective motions become more efficient in the second layer, we observe that 1) the temperature and solute gradients in the second layer become close to zero (even when the convective activity there is not as strong as in the outer convection zone, the solute and temperature are well-mixed), and 2) the interface that separates the outer convection zone and the second layer becomes narrower (see panels d-f). Additional convective layers form due to the same process. Although we present results only for one case, it is worth mentioning that we observe the same behaviour when a convective-staircase forms in the run using $\mathrm{Pr} = 7$ and $F_0/F_{\rm crit} = 10$.

\subsection{Why don’t we observe convective staircases at low Pr or low fluxes?} \label{sec:no_layers_low_pr}

\begin{figure*}
\centering
\includegraphics[width=\textwidth]{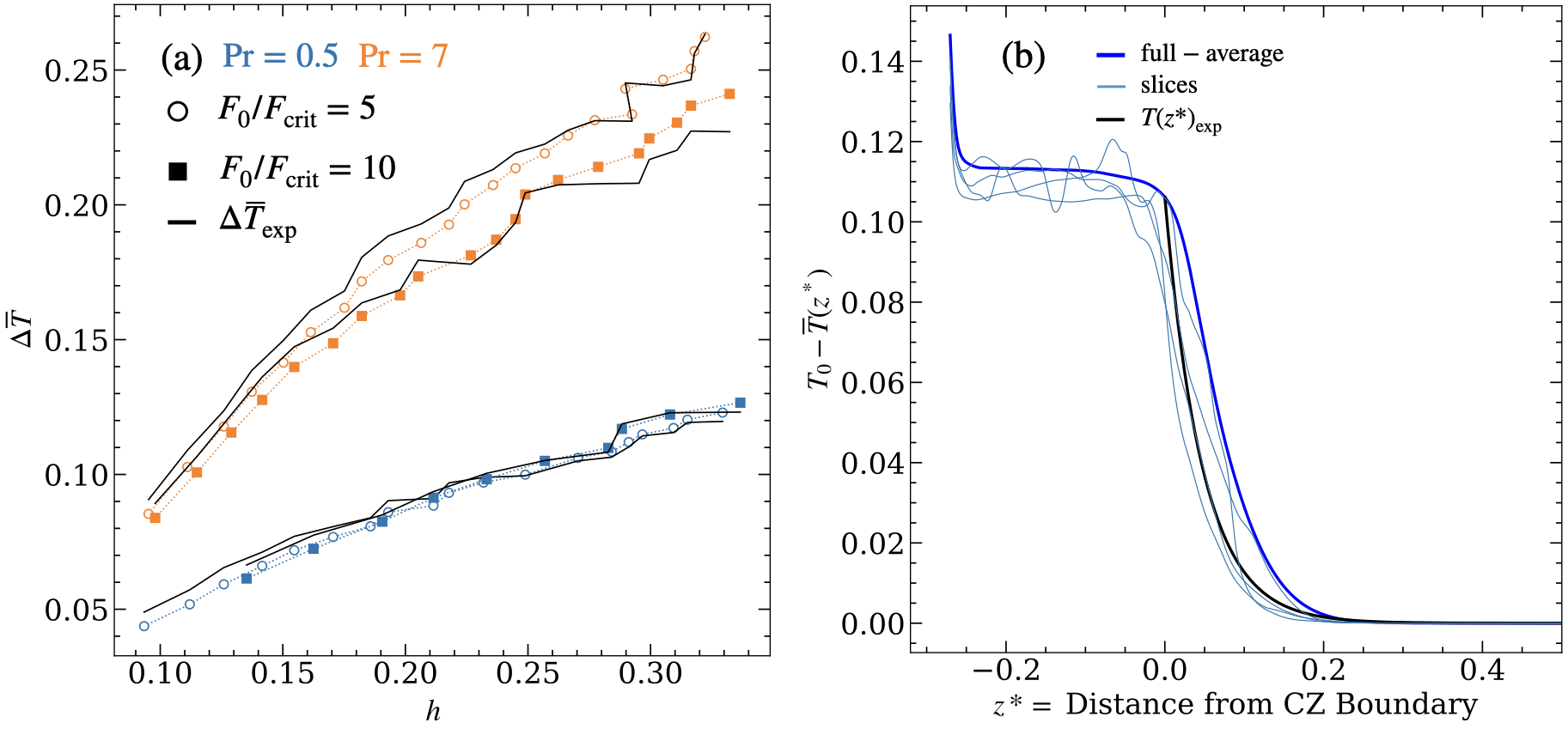}
\caption{Panel (a): Temperature step across the thermal boundary layer $\Delta \overline{T}$ as a function of the thickness of the convection zone $h$. Results are shown for simulations at low and high Pr using $F_0/F_{\mathrm{crit}} = 5$, and 10, as shown in the legends. The solid lines correspond to the expectations for $\Delta \overline{T}_{\mathrm{exp}}$ in Eq.~\eqref{eq:Delta_T_pred}, using $\varepsilon \approx 0.3$ in all the cases. Panel (b): Thermal structure (relative to $T_0$) underneath the outer convection zone. Results are shown at a particular time for the run using $\mathrm{Pr} = 0.5$ and $F_0/F_{\mathrm{crit}} = 5$. Note that $z^*$ is measured from the convective boundary, and it increases towards the bottom of the box. The black line corresponds to the expectation for $T(z^{*})_{\mathrm{exp}}$ in Eq. \eqref{eq:T_bl} using the values of $\Delta \overline{T}$ and $\dot{h}$ at the corresponding time of the profile. The blue lines are profiles measured from the simulation, using slices in the x-direction (light blue curves), and the full average over the x-direction (blue).}
\label{fig:delta_T_profile}
\end{figure*}

To understand why secondary convective layers did not form at low Pr or at low $F_0/F_{\mathrm{crit}}$, we investigate how Pr and $F_0/F_{\mathrm{crit}}$ affect the structure of the thermal boundary layer underneath the outer convection zone. We do so by calculating the temperature step at the convective boundary, $\Delta \overline{T}$, and the thickness of the thermal boundary layer, $\delta_T$, because both determine the relevant temperature gradient that competes against the stabilizing composition gradient. From horizontally-averaged profiles of the temperature field, we estimate $\Delta \overline{T}$ as the difference between the temperature at the bottom of the box, and the temperature in the convection zone. We find that at a given convection zone size $h$, $\Delta \overline{T}$ is about a factor of two larger at $\mathrm{Pr}= 7$ when compared with $\mathrm{Pr}= 0.5$ (Fig.~\ref{fig:delta_T_profile}a). This is explained by a less efficient convective transport at large Pr, meaning that it takes longer for the convection zone to reach a thickness $h$ and consequently, the temperature of the convection zone drops more than at low Pr (the fluid has cooled over longer times). Note that a larger temperature step across the boundary layer has the effect of increasing the effective temperature gradient. This is relevant since this gradient has to be large enough to overcome the composition gradient and trigger a convective instability in the boundary layer. 

It is worth mentioning that $\Delta \overline{T}$ can be obtained from conservation of the energy content within the convection zone, giving an expression for the expected temperature difference, $\Delta \overline{T}_{\mathrm\exp}$
\begin{equation}
\Delta \overline{T}_{\mathrm{exp}} =\left(\dfrac{F_0}{F_{\mathrm{crit}}}\right)\left(\dfrac{1-\varepsilon}{h}\right) t\, . \label{eq:Delta_T_pred}
\end{equation}
Using the instantaneous values of $t$, $h$, as well as $\varepsilon \approx 0.3$ \citep[measured from the heat flux at the convective boundary, for details see][]{2020PhRvF...5l4501F}, we find a reasonable agreement between the direct measurement of $\Delta \overline{T}$ and the expected value (see black lines in Fig.~\ref{fig:delta_T_profile}a). The deviations from the expectations are due to the assumption of a constant $\varepsilon$ when in reality it varies with time. Further, note that the exact dependence with $F_0/F_{\mathrm{crit}}$ relies on the form of $h(t)$. For example, for the model in Eq.~\eqref{eq_h_prediction}, $\Delta \overline{T} = h(1-\varepsilon)/2C$. However, we do not use this relation since the relation between $h$ and $t$ is steeper than 0.5.

To determine the thickness of the thermal boundary layer, we measured the temperature profile beneath the outer convection zone. \citet{turner_1968} obtained an analytical solution for it by solving the thermal diffusion equation relative to the convective boundary. In the limit where the thermal diffusion across the boundary layer is much longer than the entrainment rate $h/\dot{h}$, the solution to the thermal structure (relative to $T_0$) decays exponentially with the distance $z^*$ measured from the convective boundary
\begin{equation}
T(z^{*})_{\mathrm{exp}} = \Delta \overline{T} \exp{\left(-z^{*}/\dot{h}\right)}\, . \label{eq:T_bl}
\end{equation}
Using the instantaneous values of $\Delta \overline{T}$ in Fig.~\ref{fig:delta_T_profile}a, as well as $\dot{h}$ (measured from $h(t)$), we find that the analytic exponential profile provides a good agreement with the profiles from the simulations (being better for profiles computed from slices in the horizontal direction, see Fig.~\ref{fig:delta_T_profile}b).

\begin{figure*}
\centering
\includegraphics[width=\textwidth]{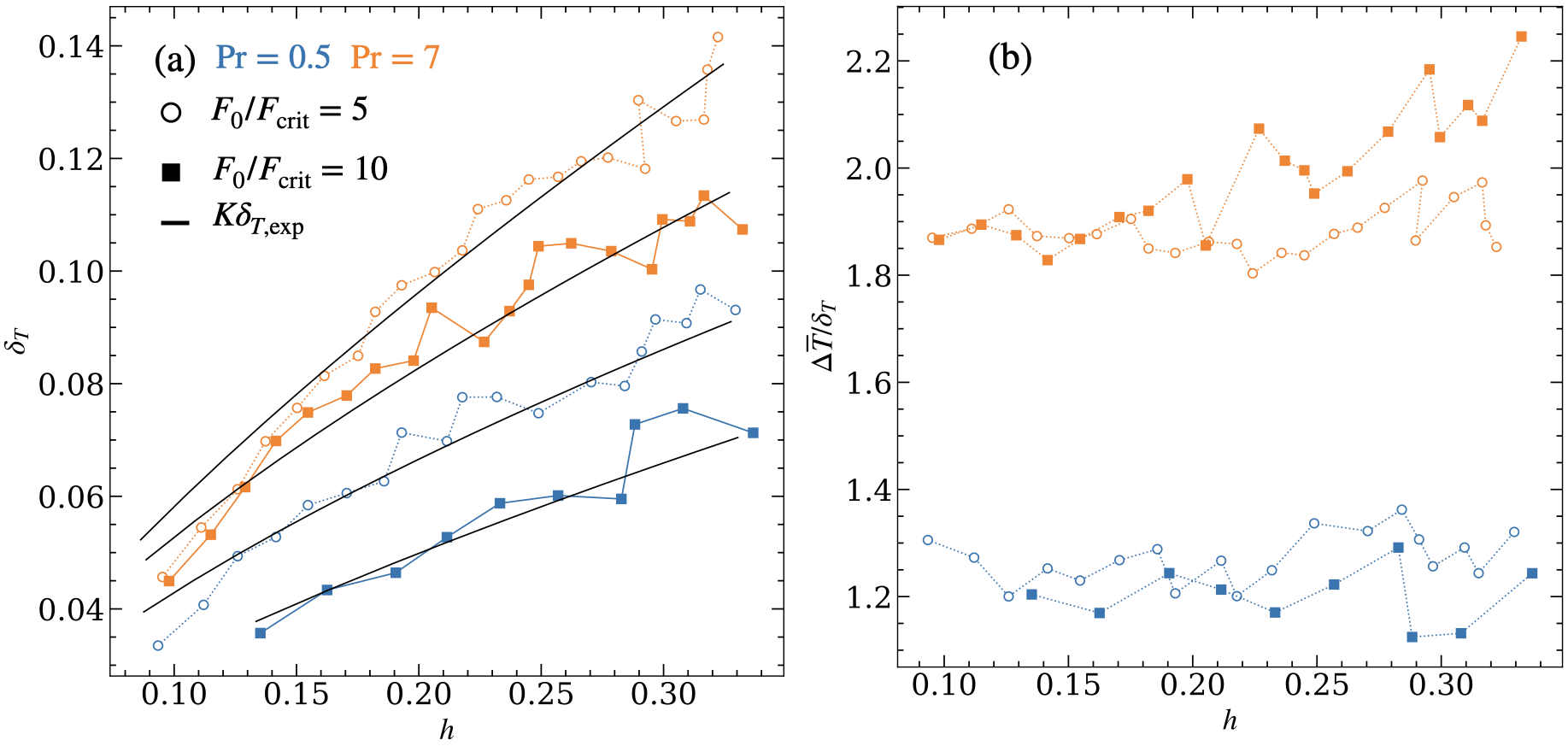}
\caption{Panel (a): Thickness of the thermal boundary layer $\delta_T$ as a function of the thickness of the convection zone $h$. The solid lines correspond to $K\delta_{T,\mathrm{exp}} = K/\dot{h}$. At $\mathrm{Pr} = 0.5$,  $K = (1.7,\, 2.4)$ for $F_0/F_{\mathrm{crit}} = (5,10)$, respectively. At $\mathrm{Pr} = 7$,  $K = (1.05,\, 2.1)$ for $F_0/F_{\mathrm{crit}} = (5,10)$, respectively. Panel (b): Ratio $\Delta\overline{T}/\delta_T$ as a function of $h$. The smaller scale for the temperature gradient at low Pr is consistent with the absence of secondary layers.}
\label{fig:delta_gradient}
\end{figure*}

Note that from Eq.~\eqref{eq:T_bl}, the expected thickness of the boundary layer is
\begin{equation}
\delta_{T,\mathrm{exp}} \sim \dfrac{1}{\dot{h}}\, \label{eq:d_bl}
\end{equation}
(where we use dimensionless variables; putting dimensions back in gives $\delta_{T,\mathrm{exp}}\sim \kappa_T/\dot{h}$).
Since we verified that Eq.~\eqref{eq:T_bl} describes well the thermal profile below the outer convection zone, we estimate $\delta_T$ by fitting the function $\Delta \overline{T}\exp(-z^{*}/\delta_T)$ to similar profiles in Fig.~\ref{fig:delta_T_profile}b. When compared with $1/\dot{h}$, we find agreement within a factor of order unity (see Fig.~\ref{fig:delta_gradient}a), in particular, $\delta_T = K\delta_{T,\mathrm{exp}}$, with $K \approx$ 1--2 being larger at low Pr and large $F_0/F_{\mathrm{crit}}$ (likely due to stronger waves at the interface, which after horizontal averaging, increase the effective width of the boundary layer). These results support the scaling in Eq.~\eqref{eq:d_bl}, i.e., at low Pr, the boundary layer is thinner because the outer convection zone moves faster than at large Pr.

\begin{figure*}
\centering
\includegraphics[width=\textwidth]{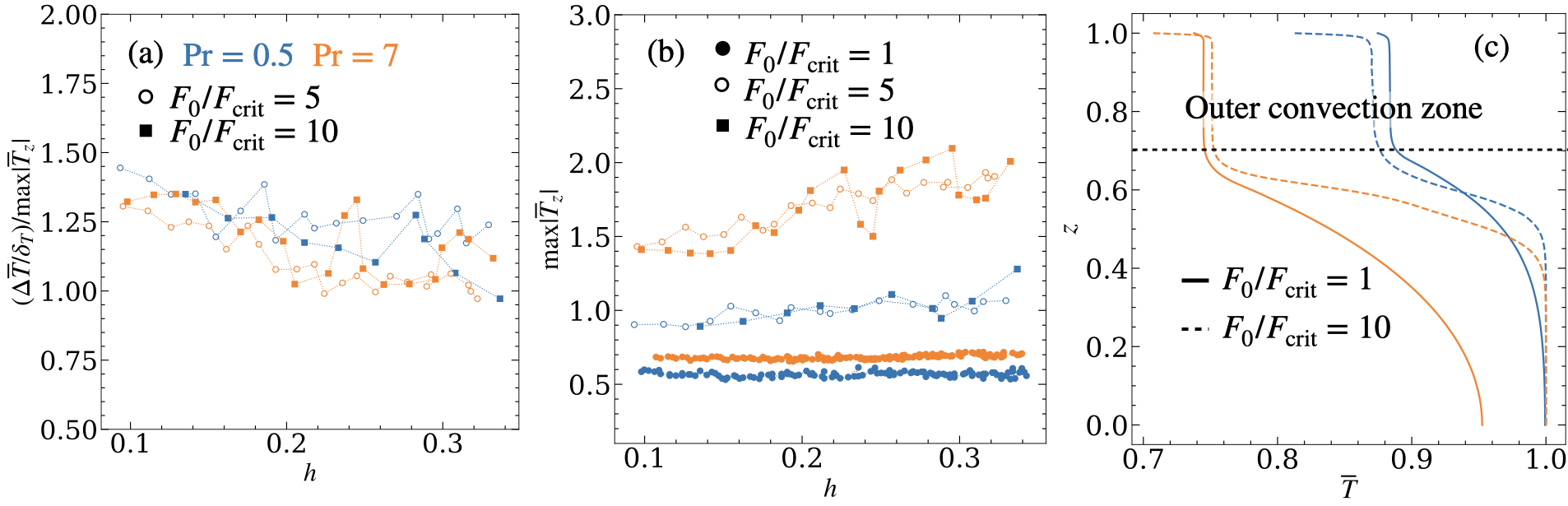}
\caption{Panel (a): Ratio $(\Delta \overline{T}/\delta_T)/\max|\overline{T}_z|$ as a function of the thickness of the convection zone $h$. Results are shown for simulations at low and high Pr using $F_0/F_{\mathrm{crit}} = 5$, and 10, as shown in the legends. Panel (b): Maximum temperature gradient within the boundary layer $\max|\overline{T}_z|$ as a function of the thickness of the convection zone $h$. Results are shown for the same simulations in panel (a) and also cases at $F_0/F_{\mathrm{crit}}=1$. Panel (c): Temperature profile for runs at low and high Pr, using $F_0/F_{\mathrm{crit}} = 1$ (solid lines) and 10 (dashed-lines). The profiles are shown at a time when the outer convection zone has advanced from the top ($z=1$) down to $z\approx 0.7$. In all panels, the blue and orange curves distinguish between simulations at $\mathrm{Pr=0.5}$ and at $\mathrm{Pr=7}$, respectively.}
\label{fig:ratio_max_Tz}
\end{figure*}

Even though the thinner boundary layer at low Pr has the effect of increasing the magnitude of the temperature gradient (becoming more unstable to convection), this is outweighed by the smaller $\Delta\overline{T}$ at low Pr, with the result that the temperature gradient is smaller at low Pr than high Pr (see Fig.~\ref{fig:delta_gradient}b). This explains why we do not observe the formation of a second layer in any of the simulations at low Pr. Note that the magnitude of $\Delta\overline{T}/\delta_T$ is $\approx 1.2$ for $\mathrm{Pr} = 0.5$, and $\approx 2$ for $\mathrm{Pr} = 7$. Since the absolute value of the composition gradient is $\approx 1$, we should expect the formation of a second layer as soon as $\Delta\overline{T}/\delta_T > 1$ (the Ledoux criterion in our dimensionless units). This is satisfied from earlier times at large Pr, but the secondary layer forms later, when $h\approx 0.35$--$4$. This is likely because $\Delta\overline{T}/\delta_T$ approximates the maximum temperature gradient within the interface (when comparing with $\max|\overline{T}_z|$ measured directly from the profiles, we find that $(\Delta\overline{T}/\delta_T)/\max|\overline{T}_z| \sim 1$--$1.4$, see Fig.~\ref{fig:ratio_max_Tz}a), whereas the second layer forms much deeper, near the inner side of the boundary layer, where the temperature gradient is much smaller. 
Figure \ref{fig:ratio_max_Tz}b shows that at low Pr the maximum temperature gradient is at most of order unity ($\lesssim 1.2$) even at the highest fluxes, consistent with the lack of secondary layers. 

A similar argument explains why we do not see the formation of a second layer at $F_0/F_{\mathrm{crit}} = 1$. In those runs, the maximum temperature gradient is much smaller than 1 due to the effects of thermal diffusion (see Fig.~\ref{fig:ratio_max_Tz}b). For runs using $F_0/F_{\mathrm{crit}} \leq 1$, the outer convection zone moves at a much slower speed, and the solution in Eq.~\eqref{eq:T_bl} is not valid \citep[see Eq.~13 in][]{turner_1968}. In fact, when looking into the structure of the thermal boundary layer underneath the outer convection zone ($z<0.7$ in Fig.~\ref{fig:ratio_max_Tz}c), we find that 1) it is much thicker than for runs using larger $F_0/F_{\mathrm{crit}}$, and 2) the temperature step at the convective boundary is also smaller than for larger $F_0/F_{\mathrm{crit}}$ runs (because the fluid has cooled down over longer times, so that the upward diffusion of heat is more significant). These two reasons explain the smaller temperature gradient for $F_0/F_{\mathrm{crit}} = 1$ in Fig.~\ref{fig:ratio_max_Tz}b).

\begin{figure*}
\centering
\includegraphics[width=\textwidth]{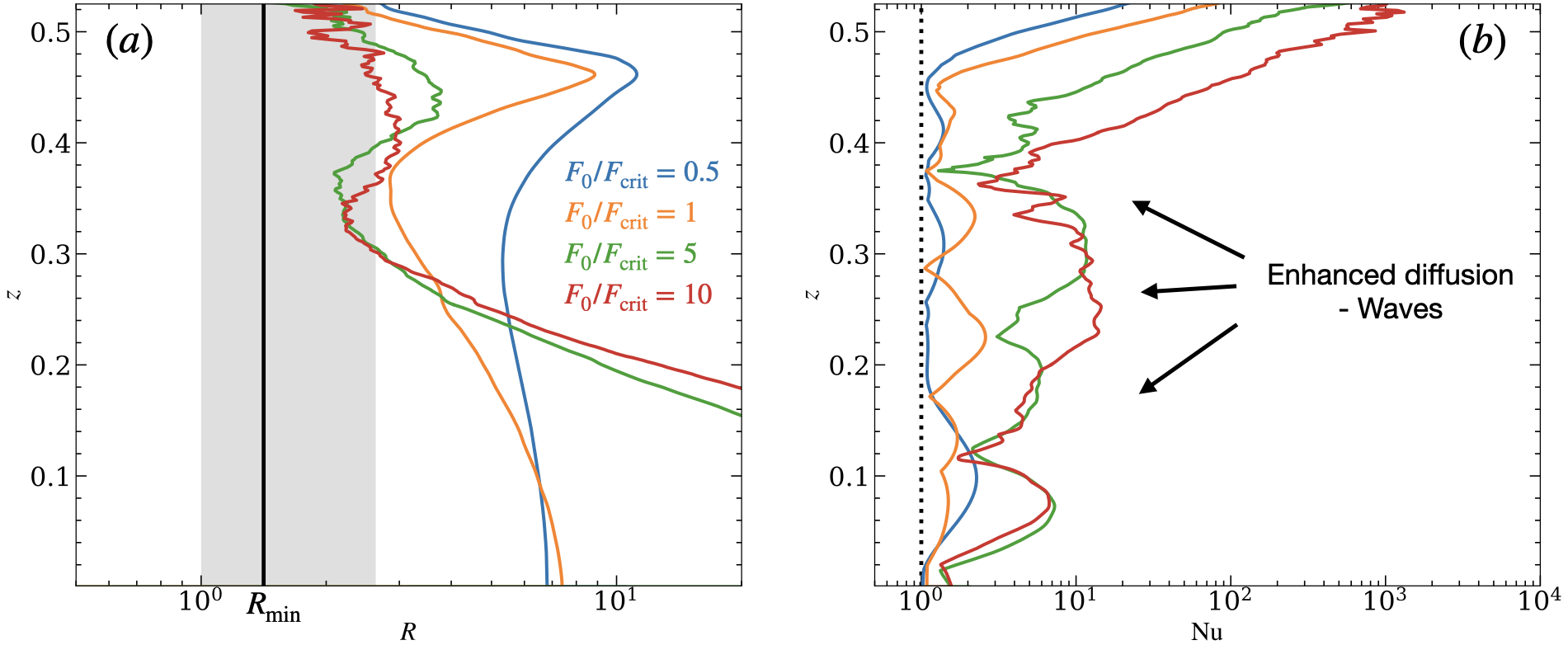}
\caption{Panel (a): Vertical profiles of the density ratio $R=S_z/T_z$ for all runs at $\mathrm{Pr}=0.5$ initialized with uniform temperature. Profiles are shown at a time when the outer convection zone has mixed all the fluid between $z = 1$ and $z=0.5$, so we exclude the upper half of the box. The gray region corresponds to the parameter space defined by $1 < R < (\mathrm{Pr} + 1)/(\mathrm{Pr} + \tau) \approx 2.6$, i.e., the possible values of $R$ for which double-diffusive instabilities are expected to occur. The black solid line corresponds to $R_{\mathrm{min}} = \mathrm{Pr}^{-1/2}\approx 1.4$ (see discussion in the text). Panel (b): Vertical profiles of the compositional Nusselt number for the same runs, time snapshots, and spatial region as in panel (a). The horizontal dotted-line corresponds to $\mathrm{Nu} = 1$, the expected value when the vertical transport is dominated by molecular diffusion.} \label{fig_R_Nu}
\end{figure*}

Even though the thermal boundary layer is stable by the Ledoux criterion at low Pr, there is the possibility of forming layers by double-diffusive instabilities. These instabilities are expected to occur when the density ratio, $R=S_z/T_z$\footnote{We have used dimensionless variables; putting dimensions back $R =\beta S_z/\alpha T_z$.}, lies in the range $1< R < (\mathrm{Pr} + 1)/(\mathrm{Pr} + \tau)$ \citep[e.g][]{2012ApJ...750...61M,2013ApJ...768..157W,2018AnRFM..50..275G}. However, it is likely that layer formation occurs only in a narrower range $1 < R < R_{\mathrm{min}}$, where $R_\mathrm{min}$ is uncertain but has been estimated to be $R_{\mathrm{min}}\approx \mathrm{Pr}^{-1/2}$ \citep{2012ApJ...750...61M,2013ApJ...768..157W}. For $R > R_{\mathrm{min}} $, the fluid evolves into a state of enhanced diffusion instead of forming layers \citep{2012ApJ...750...61M, 2013ApJ...768..157W}. In this work we have Pr$=0.5$, and $\tau = 0.07$, giving the upper bound for double-diffusive instabilities as $(\mathrm{Pr} + 1)/(\mathrm{Pr} + \tau) \approx 2.6$, and $R_{\mathrm{min}} \approx 1.4$. 
Interestingly, we find that the density ratio in our low Pr simulations does enter the range where double-diffusive instabilities are expected to occur for the larger values of $F_0/F_\mathrm{crit}$. However, $R$ is always larger than $R_\mathrm{min}$, so the fluid is in the regime of enhanced diffusion. This is shown in Fig.~\ref{fig_R_Nu} where we show profiles of $R$, and composition Nusselt number Nu$\equiv 1 + \overline{wS}/\tau|\overline{S}_z|$ for simulations with $\mathrm{Pr}=0.5$ and different values of $F_0/F_{\mathrm{crit}}$. The Nusselt number is clearly larger in the simulations for which $R$ enters the double-diffusive unstable region (gray region). So although double-diffusive instabilities do not lead to layer formation, they do appear to play a role in the region below the outer convection zone.

\begin{figure*}
\centering
\includegraphics[width=0.8\textwidth]{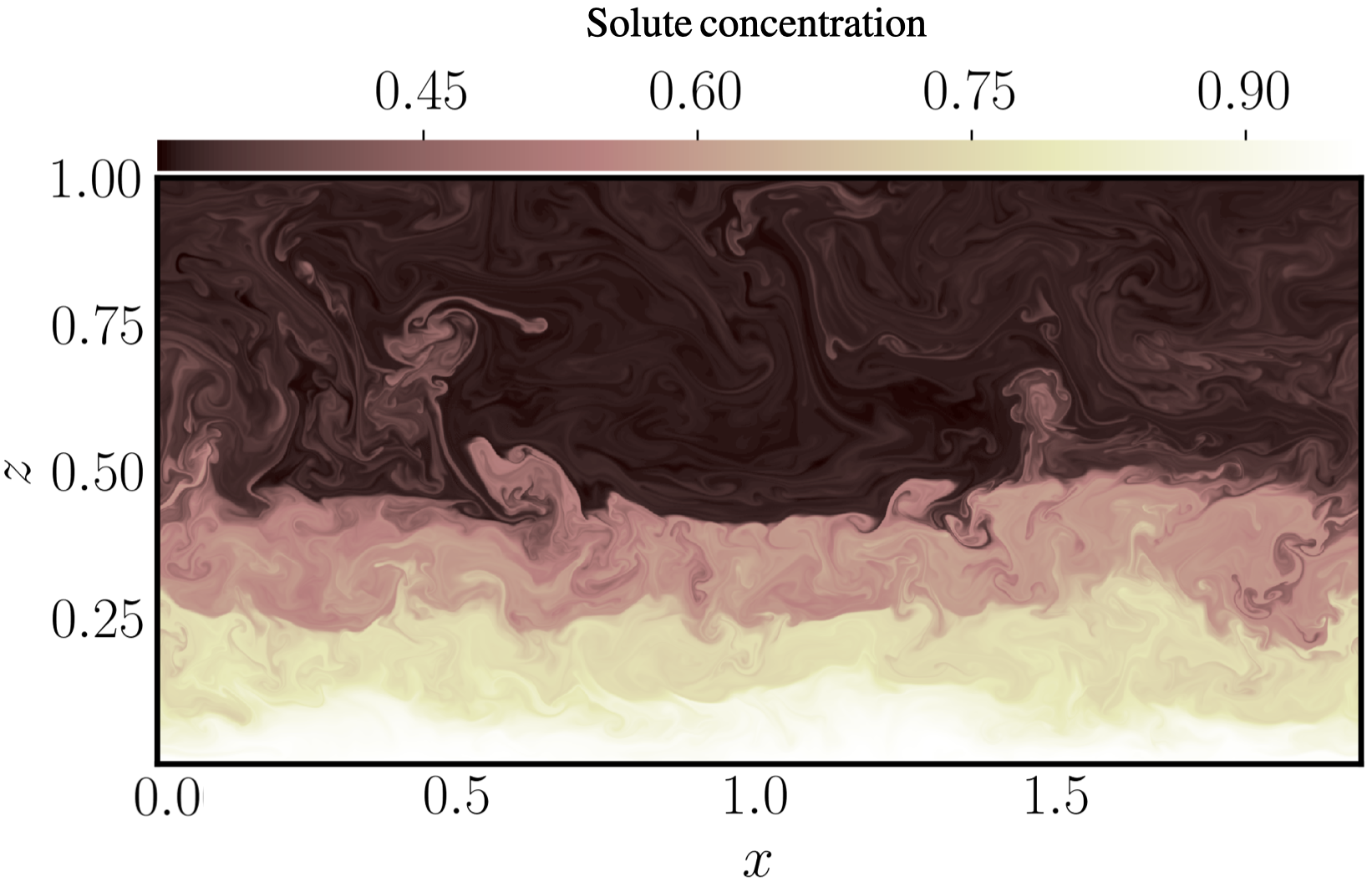}
\caption{2D snapshots of the solute field for a run using $\mathrm{Pr}=0.5$, $F_0/F_{\rm crit}=5$, $dS_0/dz = -1$, and $dT_0/dz = -0.8$ (so that $R_0 = 1.25$). The snapshot is shown at a particular time when additional convective layers are clearly visible. Unlike the layers at large Pr in Fig. \ref{fig_S_layers}, layers due to double-diffusive instabilities form spontaneously. }\label{fig_S_R}
\end{figure*}

Layered convection and convective-staircases have been found at low Pr in numerical experiments which are initialized with appropriate background gradients of temperature and solute \citep[such that double-diffusive instabilities can be triggered, see review by ][]{2018AnRFM..50..275G}. 
We conducted additional experiments with linear temperature profiles in the range where layers are expected, to confirm that they do develop in our time-dependent problem. Based on the value of $R_\mathrm{min}$ above, this occurs only in the narrow range of initial temperature gradients $-1 < dT_0/dz \lesssim -0.71$ for our values of Pr and $\tau$.
An example is shown in Fig.~\ref{fig_S_R}. In this case, because we have chosen the gradients appropriately, layers spontaneously develop throughout the box due to double-diffusive instabilities. However, we find that they are not long-lived because they subsequently are engulfed by the outer convection zone as it propagates inwards, and once again the box is fully-mixed by the end of the simulation.

\section{Summary and Conclusions} \label{sec:conclusions} 

Motivated by evolutionary models of Jupiter that show the formation of multiple long-lived convective layers, we studied a fluid with a stable composition gradient that is constantly cooled from the top. As soon as the cooling flux is activated at the top boundary, a steep temperature gradient develops and forms an outer convection zone that grows inwards by entrainment of heavier fluid from below. We performed simulations at $\mathrm{Pr} = 0.5$ and 7, varying the magnitude of the cooling flux. Our goal was to test whether secondary convective layers form below the outer convection zone. In summary:

\begin{enumerate}

\item At large Pr, we find multiple convective layers form as long as the heat flux driving convection is sufficiently large. These layers develop over time and their formation resembles the dynamics of layers in laboratory experiments, i.e.~instability of the thermal boundary layer below the outer convection zone once the Ledoux criterion is satisfied. These layers persist for a short time until they get entrained into the outer convection zone. 

\item In contrast to high Pr, layers do not form at low Pr, and the outer convection zone completely mixes the primordial composition gradient, no matter the magnitude of the cooling flux $F_0/F_{\rm crit}$. This difference is explained by the effect of Pr on the structure of the thermal boundary layer underneath the outer convection zone. Since at low Pr the convective transport is much more efficient than at large Pr, both the thickness of the boundary layer and temperature step across it are much smaller than at large Pr. In particular, both quantities evolve in such a way that the temperature gradient underneath the outer convection zone is much smaller than at large Pr. Consequently, the destabilizing effect of the thermal stratification is not large enough to overcome the solute gradient and trigger convective instabilities to form a staircase.

\end{enumerate}

We do not rule out the possibility that at low Pr, the thermal boundary layer below the outer convection zone could become convectively unstable. Experiments using a wider range of parameters (mainly in Pr, $\mathrm{\tau}$, and $F_0/F_{\rm crit}$) and different compositional stratifications are needed to verify this. Recent numerical simulations carried out by \citet{2019ThCFD..33..383Z} have shown the formation of multiple convective layers at low Pr in a time-dependent situation. The numerical setup in \citet{2019ThCFD..33..383Z} differs from this work mainly in the boundary conditions. The temperature and solute are fixed at the boundaries, giving prescribed average gradients across the box. Further, the main convection zone is driven by a heat flux that decreases over time. From the information presented there, we estimate that $F_0/F_{\rm crit}$ decreases over time from $\approx 16$ to 1. These values of the cooling flux are similar to the ones used in our work. Interestingly, they found that layers form by two different mechanisms: a second layer due to a convective instability in the thermal boundary layer ahead of the front, and the spontaneous formation of multiple layers due to double-diffusive instabilities. Our experiments did not exhibit this behaviour, presumably because the initial temperature was constant everywhere within the box, and the temperature gradient develops over time as the fluid cools from the top boundary. Another difference with \citet{2019ThCFD..33..383Z} is that in their setup the outer convection zone stalls (likely due to the decreasing flux that drives convection). In our simulations the outer convection zone continues to penetrate inwards and no secondary layer forms at low Pr. Simulations with a time-dependent heat flux at the top boundary, e.g. $\partial T/\partial z \propto T(H,t)$, would be useful to verify under what conditions the outer convective layer stops propagating inward over long timescales.

Our results suggest that below an evolving convection zone, the formation of layers is more difficult at low Pr. This may have implications for the ability of composition gradients to survive in Jupiter’s interior. For example, in 1D evolutionary models of Jupiter, the initial composition gradient forms layers that survive until the present day \citep{2018A&A...610L..14V,2020A&A...638A.121M,2022PSJ.....3...74S}. This is important for interpreting the Juno data that suggest an extended, dilute core in Jupiter \cite{2017GeoRL..44.4649W}. From the experiments here, entrainment of heavy elements into the outer convection zone may actually prevent the formation of such a staircase. Yet, we stress that our model is far from being representative of the conditions in Jupiter's interior. While our simulations include multi-dimensional convective turbulence, they are idealized compared to 1D evolution models in many aspects. For example, we did not model the entire planet, and did not use a complex equation of state for mixtures. Also, our simulations consider constant diffusivities. Also, we restricted our simulations to fluids moving in 2D, while in the real planet the fluid motions are 3D. This simplication could change the results in this work since previous studies of thermal convection have shown that at low Pr, the convective transport in 2D simulations is much larger than in 3D \citep[e.g.,][]{2004EL.....67..390S}. This affects the structure of the thermal boundary layer responsible for the formation of the second convective layer, as the temperature step across it would be larger (increasing the temperature gradient). Also, we ignore the fact that density varies by many orders of magnitude in Jupiter's interior. Accounting for compressibility effects could also affect the transport properties. This is because the density increases over many scales as the fluid moves from the surface toward the interior. Therefore, the fluid velocities should be smaller deeper in the fluid. This does not happen in Boussinesq flows, since the background density is roughly constant. Although simulations at low Pr and low $\tau$ are possible, rough estimations from primordial profiles of Jupiter (Simon Müller, personal communication) give $F_0/F_{\rm crit} \sim 1$--$1000$, somewhat larger than considered here, whereas the Rayleigh number based on the planet radius is significantly larger, $\mathcal{R} \sim 10^{37}$. It is also important to note that gas giant planets do not cool over time at a constant rate. The luminosity of the planet decreases over time, and so does the strength of the heat flux that drives the evolution of the outer convection zone.  Finally, in Jupiter the convective turnover time is significantly larger than its rotation period (the Rossby number is $\sim 10^{-5}$--$10^{-4}$). Therefore, the convective dynamics is highly constrained by rotation. To the extent possible, future work should approach this problem considering spherical geometry, rotation, density stratification, and a time-dependent cooling flux.
Also, a primordial distribution of solute with a larger concentration near the center would be more appropriate \citep[as predicted from recent formation models,  e.g.][]{2022PSJ.....3...74S}. This will be crucial to improve our understanding of Jupiter's interior, as well as to explain observations of Jupiter and other gas giants.

\begin{acknowledgements}
We thank the anonymous referees for providing a careful report that helped to improve the manuscript. This work was supported by an NSERC Discovery Grant. J.~R.~F. acknowledges support from a McGill Space Institute (MSI) Fellowship, and thanks the Department of Applied Mathematics at the University of Colorado Boulder, for hospitality. We thank Simon M\"uller for providing primordial profiles of Jupiter, and Tristan Guillot, Ravit Helled and Allona Vazan for insightful discussions about layer formation in gas giant models. We are grateful to D. J. Stevenson for pointing out useful references. E.~H.~A.~is supported by CIERA and Northwestern University through a CIERA Postdoctoral fellowship. A.~C.~and J.~R.~F.~are members of the Centre de Recherche en Astrophysique du Québec (CRAQ) and the Institut de recherche sur les exoplanètes (iREx). This research was enabled in part by support provided by Calcul Québec (calculquebec.ca), and Compute Canada (www.computecanada.ca). Computations were performed on Graham and Béluga.
\end{acknowledgements}

\bibliography{references}

\begin{thebibliography}{36}%
\makeatletter
\providecommand \@ifxundefined [1]{%
 \@ifx{#1\undefined}
}%
\providecommand \@ifnum [1]{%
 \ifnum #1\expandafter \@firstoftwo
 \else \expandafter \@secondoftwo
 \fi
}%
\providecommand \@ifx [1]{%
 \ifx #1\expandafter \@firstoftwo
 \else \expandafter \@secondoftwo
 \fi
}%
\providecommand \natexlab [1]{#1}%
\providecommand \enquote  [1]{``#1''}%
\providecommand \bibnamefont  [1]{#1}%
\providecommand \bibfnamefont [1]{#1}%
\providecommand \citenamefont [1]{#1}%
\providecommand \href@noop [0]{\@secondoftwo}%
\providecommand \href [0]{\begingroup \@sanitize@url \@href}%
\providecommand \@href[1]{\@@startlink{#1}\@@href}%
\providecommand \@@href[1]{\endgroup#1\@@endlink}%
\providecommand \@sanitize@url [0]{\catcode `\\12\catcode `\$12\catcode
  `\&12\catcode `\#12\catcode `\^12\catcode `\_12\catcode `\%12\relax}%
\providecommand \@@startlink[1]{}%
\providecommand \@@endlink[0]{}%
\providecommand \url  [0]{\begingroup\@sanitize@url \@url }%
\providecommand \@url [1]{\endgroup\@href {#1}{\urlprefix }}%
\providecommand \urlprefix  [0]{URL }%
\providecommand \Eprint [0]{\href }%
\providecommand \doibase [0]{https://doi.org/}%
\providecommand \selectlanguage [0]{\@gobble}%
\providecommand \bibinfo  [0]{\@secondoftwo}%
\providecommand \bibfield  [0]{\@secondoftwo}%
\providecommand \translation [1]{[#1]}%
\providecommand \BibitemOpen [0]{}%
\providecommand \bibitemStop [0]{}%
\providecommand \bibitemNoStop [0]{.\EOS\space}%
\providecommand \EOS [0]{\spacefactor3000\relax}%
\providecommand \BibitemShut  [1]{\csname bibitem#1\endcsname}%
\let\auto@bib@innerbib\@empty
\bibitem [{\citenamefont {{Bolton}}\ \emph {et~al.}(2017)\citenamefont
  {{Bolton}}, \citenamefont {{Lunine}}, \citenamefont {{Stevenson}},
  \citenamefont {{Connerney}}, \citenamefont {{Levin}}, \citenamefont {{Owen}},
  \citenamefont {{Bagenal}}, \citenamefont {{Gautier}}, \citenamefont
  {{Ingersoll}}, \citenamefont {{Orton}}, \citenamefont {{Guillot}},
  \citenamefont {{Hubbard}}, \citenamefont {{Bloxham}}, \citenamefont
  {{Coradini}}, \citenamefont {{Stephens}}, \citenamefont {{Mokashi}},
  \citenamefont {{Thorne}},\ and\ \citenamefont
  {{Thorpe}}}]{2017SSRv..213....5B}%
  \BibitemOpen
  \bibfield  {author} {\bibinfo {author} {\bibfnamefont {S.~J.}\ \bibnamefont
  {{Bolton}}}, \bibinfo {author} {\bibfnamefont {J.}~\bibnamefont {{Lunine}}},
  \bibinfo {author} {\bibfnamefont {D.}~\bibnamefont {{Stevenson}}}, \bibinfo
  {author} {\bibfnamefont {J.~E.~P.}\ \bibnamefont {{Connerney}}}, \bibinfo
  {author} {\bibfnamefont {S.}~\bibnamefont {{Levin}}}, \bibinfo {author}
  {\bibfnamefont {T.~C.}\ \bibnamefont {{Owen}}}, \bibinfo {author}
  {\bibfnamefont {F.}~\bibnamefont {{Bagenal}}}, \bibinfo {author}
  {\bibfnamefont {D.}~\bibnamefont {{Gautier}}}, \bibinfo {author}
  {\bibfnamefont {A.~P.}\ \bibnamefont {{Ingersoll}}}, \bibinfo {author}
  {\bibfnamefont {G.~S.}\ \bibnamefont {{Orton}}}, \bibinfo {author}
  {\bibfnamefont {T.}~\bibnamefont {{Guillot}}}, \bibinfo {author}
  {\bibfnamefont {W.}~\bibnamefont {{Hubbard}}}, \bibinfo {author}
  {\bibfnamefont {J.}~\bibnamefont {{Bloxham}}}, \bibinfo {author}
  {\bibfnamefont {A.}~\bibnamefont {{Coradini}}}, \bibinfo {author}
  {\bibfnamefont {S.~K.}\ \bibnamefont {{Stephens}}}, \bibinfo {author}
  {\bibfnamefont {P.}~\bibnamefont {{Mokashi}}}, \bibinfo {author}
  {\bibfnamefont {R.}~\bibnamefont {{Thorne}}},\ and\ \bibinfo {author}
  {\bibfnamefont {R.}~\bibnamefont {{Thorpe}}},\ }\bibfield  {title} {\bibinfo
  {title} {{The Juno Mission}},\ }\href
  {https://doi.org/10.1007/s11214-017-0429-6} {\bibfield  {journal} {\bibinfo
  {journal} {Space Sci. Rev.}\ }\textbf {\bibinfo {volume} {213}},\ \bibinfo
  {pages} {5} (\bibinfo {year} {2017})}\BibitemShut {NoStop}%
\bibitem [{\citenamefont {{Wahl}}\ \emph {et~al.}(2017)\citenamefont {{Wahl}},
  \citenamefont {{Hubbard}}, \citenamefont {{Militzer}}, \citenamefont
  {{Guillot}}, \citenamefont {{Miguel}}, \citenamefont {{Movshovitz}},
  \citenamefont {{Kaspi}}, \citenamefont {{Helled}}, \citenamefont {{Reese}},
  \citenamefont {{Galanti}}, \citenamefont {{Levin}}, \citenamefont
  {{Connerney}},\ and\ \citenamefont {{Bolton}}}]{2017GeoRL..44.4649W}%
  \BibitemOpen
  \bibfield  {author} {\bibinfo {author} {\bibfnamefont {S.~M.}\ \bibnamefont
  {{Wahl}}}, \bibinfo {author} {\bibfnamefont {W.~B.}\ \bibnamefont
  {{Hubbard}}}, \bibinfo {author} {\bibfnamefont {B.}~\bibnamefont
  {{Militzer}}}, \bibinfo {author} {\bibfnamefont {T.}~\bibnamefont
  {{Guillot}}}, \bibinfo {author} {\bibfnamefont {Y.}~\bibnamefont {{Miguel}}},
  \bibinfo {author} {\bibfnamefont {N.}~\bibnamefont {{Movshovitz}}}, \bibinfo
  {author} {\bibfnamefont {Y.}~\bibnamefont {{Kaspi}}}, \bibinfo {author}
  {\bibfnamefont {R.}~\bibnamefont {{Helled}}}, \bibinfo {author}
  {\bibfnamefont {D.}~\bibnamefont {{Reese}}}, \bibinfo {author} {\bibfnamefont
  {E.}~\bibnamefont {{Galanti}}}, \bibinfo {author} {\bibfnamefont
  {S.}~\bibnamefont {{Levin}}}, \bibinfo {author} {\bibfnamefont {J.~E.}\
  \bibnamefont {{Connerney}}},\ and\ \bibinfo {author} {\bibfnamefont {S.~J.}\
  \bibnamefont {{Bolton}}},\ }\bibfield  {title} {\bibinfo {title} {{Comparing
  Jupiter interior structure models to Juno gravity measurements and the role
  of a dilute core}},\ }\href {https://doi.org/10.1002/2017GL073160} {\bibfield
   {journal} {\bibinfo  {journal} {Geophys. Res. Lett.}\ }\textbf {\bibinfo
  {volume} {44}},\ \bibinfo {pages} {4649} (\bibinfo {year}
  {2017})}\BibitemShut {NoStop}%
\bibitem [{\citenamefont {{Ingersoll}}(2020)}]{2020SSRv..216..122I}%
  \BibitemOpen
  \bibfield  {author} {\bibinfo {author} {\bibfnamefont {A.~P.}\ \bibnamefont
  {{Ingersoll}}},\ }\bibfield  {title} {\bibinfo {title} {{Cassini Exploration
  of the Planet Saturn: A Comprehensive Review}},\ }\href
  {https://doi.org/10.1007/s11214-020-00751-1} {\bibfield  {journal} {\bibinfo
  {journal} {Space Sci. Rev.}\ }\textbf {\bibinfo {volume} {216}},\ \bibinfo
  {eid} {122} (\bibinfo {year} {2020})}\BibitemShut {NoStop}%
\bibitem [{\citenamefont {{Mankovich}}\ and\ \citenamefont
  {{Fuller}}(2021)}]{2021NatAs...5.1103M}%
  \BibitemOpen
  \bibfield  {author} {\bibinfo {author} {\bibfnamefont {C.~R.}\ \bibnamefont
  {{Mankovich}}}\ and\ \bibinfo {author} {\bibfnamefont {J.}~\bibnamefont
  {{Fuller}}},\ }\bibfield  {title} {\bibinfo {title} {{A diffuse core in
  Saturn revealed by ring seismology}},\ }\href
  {https://doi.org/10.1038/s41550-021-01448-3} {\bibfield  {journal} {\bibinfo
  {journal} {Nature Astronomy}\ }\textbf {\bibinfo {volume} {5}},\ \bibinfo
  {pages} {1103} (\bibinfo {year} {2021})}\BibitemShut {NoStop}%
\bibitem [{\citenamefont {{Miguel}}\ \emph {et~al.}(2016)\citenamefont
  {{Miguel}}, \citenamefont {{Guillot}},\ and\ \citenamefont
  {{Fayon}}}]{2016A&A...596A.114M}%
  \BibitemOpen
  \bibfield  {author} {\bibinfo {author} {\bibfnamefont {Y.}~\bibnamefont
  {{Miguel}}}, \bibinfo {author} {\bibfnamefont {T.}~\bibnamefont
  {{Guillot}}},\ and\ \bibinfo {author} {\bibfnamefont {L.}~\bibnamefont
  {{Fayon}}},\ }\bibfield  {title} {\bibinfo {title} {{Jupiter internal
  structure: the effect of different equations of state}},\ }\href
  {https://doi.org/10.1051/0004-6361/201629732} {\bibfield  {journal} {\bibinfo
   {journal} {Astron. Astrophys.}\ }\textbf {\bibinfo {volume} {596}},\
  \bibinfo {eid} {A114} (\bibinfo {year} {2016})}\BibitemShut {NoStop}%
\bibitem [{\citenamefont {{Stevenson}}(1985)}]{1985Icar...62....4S}%
  \BibitemOpen
  \bibfield  {author} {\bibinfo {author} {\bibfnamefont {D.~J.}\ \bibnamefont
  {{Stevenson}}},\ }\bibfield  {title} {\bibinfo {title} {{Cosmochemistry and
  structure of the giant planets and their satellites}},\ }\href
  {https://doi.org/10.1016/0019-1035(85)90168-X} {\bibfield  {journal}
  {\bibinfo  {journal} {Icar}\ }\textbf {\bibinfo {volume} {62}},\ \bibinfo
  {pages} {4} (\bibinfo {year} {1985})}\BibitemShut {NoStop}%
\bibitem [{\citenamefont {{Leconte}}\ and\ \citenamefont
  {{Chabrier}}(2012)}]{2012A&A...540A..20L}%
  \BibitemOpen
  \bibfield  {author} {\bibinfo {author} {\bibfnamefont {J.}~\bibnamefont
  {{Leconte}}}\ and\ \bibinfo {author} {\bibfnamefont {G.}~\bibnamefont
  {{Chabrier}}},\ }\bibfield  {title} {\bibinfo {title} {{A new vision of giant
  planet interiors: Impact of double diffusive convection}},\ }\href
  {https://doi.org/10.1051/0004-6361/201117595} {\bibfield  {journal} {\bibinfo
   {journal} {Astron. Astrophys.}\ }\textbf {\bibinfo {volume} {540}},\
  \bibinfo {eid} {A20} (\bibinfo {year} {2012})}\BibitemShut {NoStop}%
\bibitem [{\citenamefont {{Garaud}}(2018)}]{2018AnRFM..50..275G}%
  \BibitemOpen
  \bibfield  {author} {\bibinfo {author} {\bibfnamefont {P.}~\bibnamefont
  {{Garaud}}},\ }\bibfield  {title} {\bibinfo {title} {{Double-Diffusive
  Convection at Low Prandtl Number}},\ }\href
  {https://doi.org/10.1146/annurev-fluid-122316-045234} {\bibfield  {journal}
  {\bibinfo  {journal} {Annu. Rev. Fluid Mech.}\ }\textbf {\bibinfo {volume}
  {50}},\ \bibinfo {pages} {275} (\bibinfo {year} {2018})}\BibitemShut
  {NoStop}%
\bibitem [{\citenamefont {{Moll}}\ \emph {et~al.}(2017)\citenamefont {{Moll}},
  \citenamefont {{Garaud}}, \citenamefont {{Mankovich}},\ and\ \citenamefont
  {{Fortney}}}]{2017ApJ...849...24M}%
  \BibitemOpen
  \bibfield  {author} {\bibinfo {author} {\bibfnamefont {R.}~\bibnamefont
  {{Moll}}}, \bibinfo {author} {\bibfnamefont {P.}~\bibnamefont {{Garaud}}},
  \bibinfo {author} {\bibfnamefont {C.}~\bibnamefont {{Mankovich}}},\ and\
  \bibinfo {author} {\bibfnamefont {J.~J.}\ \bibnamefont {{Fortney}}},\
  }\bibfield  {title} {\bibinfo {title} {{Double-diffusive Erosion of the Core
  of Jupiter}},\ }\href {https://doi.org/10.3847/1538-4357/aa8d74} {\bibfield
  {journal} {\bibinfo  {journal} {Astrophys. J.}\ }\textbf {\bibinfo {volume}
  {849}},\ \bibinfo {eid} {24} (\bibinfo {year} {2017})}\BibitemShut {NoStop}%
\bibitem [{\citenamefont {{Chabrier}}\ and\ \citenamefont
  {{Baraffe}}(2007)}]{2007ApJ...661L..81C}%
  \BibitemOpen
  \bibfield  {author} {\bibinfo {author} {\bibfnamefont {G.}~\bibnamefont
  {{Chabrier}}}\ and\ \bibinfo {author} {\bibfnamefont {I.}~\bibnamefont
  {{Baraffe}}},\ }\bibfield  {title} {\bibinfo {title} {{Heat Transport in
  Giant (Exo)planets: A New Perspective}},\ }\href
  {https://doi.org/10.1086/518473} {\bibfield  {journal} {\bibinfo  {journal}
  {Astrophys. J. Lett.}\ }\textbf {\bibinfo {volume} {661}},\ \bibinfo {pages}
  {L81} (\bibinfo {year} {2007})}\BibitemShut {NoStop}%
\bibitem [{\citenamefont {{Leconte}}\ and\ \citenamefont
  {{Chabrier}}(2013)}]{2013NatGe...6..347L}%
  \BibitemOpen
  \bibfield  {author} {\bibinfo {author} {\bibfnamefont {J.}~\bibnamefont
  {{Leconte}}}\ and\ \bibinfo {author} {\bibfnamefont {G.}~\bibnamefont
  {{Chabrier}}},\ }\bibfield  {title} {\bibinfo {title} {{Layered convection as
  the origin of Saturn's luminosity anomaly}},\ }\href
  {https://doi.org/10.1038/ngeo1791} {\bibfield  {journal} {\bibinfo  {journal}
  {NatGe}\ }\textbf {\bibinfo {volume} {6}},\ \bibinfo {pages} {347} (\bibinfo
  {year} {2013})}\BibitemShut {NoStop}%
\bibitem [{\citenamefont {{Vazan}}\ \emph {et~al.}(2018)\citenamefont
  {{Vazan}}, \citenamefont {{Helled}},\ and\ \citenamefont
  {{Guillot}}}]{2018A&A...610L..14V}%
  \BibitemOpen
  \bibfield  {author} {\bibinfo {author} {\bibfnamefont {A.}~\bibnamefont
  {{Vazan}}}, \bibinfo {author} {\bibfnamefont {R.}~\bibnamefont {{Helled}}},\
  and\ \bibinfo {author} {\bibfnamefont {T.}~\bibnamefont {{Guillot}}},\
  }\bibfield  {title} {\bibinfo {title} {{Jupiter's evolution with primordial
  composition gradients}},\ }\href
  {https://doi.org/10.1051/0004-6361/201732522} {\bibfield  {journal} {\bibinfo
   {journal} {Astron. Astrophys.}\ }\textbf {\bibinfo {volume} {610}},\
  \bibinfo {eid} {L14} (\bibinfo {year} {2018})}\BibitemShut {NoStop}%
\bibitem [{\citenamefont {{M{\"u}ller}}\ \emph {et~al.}(2020)\citenamefont
  {{M{\"u}ller}}, \citenamefont {{Helled}},\ and\ \citenamefont
  {{Cumming}}}]{2020A&A...638A.121M}%
  \BibitemOpen
  \bibfield  {author} {\bibinfo {author} {\bibfnamefont {S.}~\bibnamefont
  {{M{\"u}ller}}}, \bibinfo {author} {\bibfnamefont {R.}~\bibnamefont
  {{Helled}}},\ and\ \bibinfo {author} {\bibfnamefont {A.}~\bibnamefont
  {{Cumming}}},\ }\bibfield  {title} {\bibinfo {title} {{The challenge of
  forming a fuzzy core in Jupiter}},\ }\href
  {https://doi.org/10.1051/0004-6361/201937376} {\bibfield  {journal} {\bibinfo
   {journal} {Astron. Astrophys.}\ }\textbf {\bibinfo {volume} {638}},\
  \bibinfo {eid} {A121} (\bibinfo {year} {2020})}\BibitemShut {NoStop}%
\bibitem [{\citenamefont {{Stevenson}}\ \emph {et~al.}(2022)\citenamefont
  {{Stevenson}}, \citenamefont {{Bodenheimer}}, \citenamefont {{Lissauer}},\
  and\ \citenamefont {{D'Angelo}}}]{2022PSJ.....3...74S}%
  \BibitemOpen
  \bibfield  {author} {\bibinfo {author} {\bibfnamefont {D.~J.}\ \bibnamefont
  {{Stevenson}}}, \bibinfo {author} {\bibfnamefont {P.}~\bibnamefont
  {{Bodenheimer}}}, \bibinfo {author} {\bibfnamefont {J.~J.}\ \bibnamefont
  {{Lissauer}}},\ and\ \bibinfo {author} {\bibfnamefont {G.}~\bibnamefont
  {{D'Angelo}}},\ }\bibfield  {title} {\bibinfo {title} {{Mixing of Condensable
  Constituents with H-He during the Formation and Evolution of Jupiter}},\
  }\href {https://doi.org/10.3847/PSJ/ac5c44} {\bibfield  {journal} {\bibinfo
  {journal} {Planet. Sci. J.}\ }\textbf {\bibinfo {volume} {3}},\ \bibinfo
  {eid} {74} (\bibinfo {year} {2022})}\BibitemShut {NoStop}%
\bibitem [{\citenamefont {{Stevenson}}\ and\ \citenamefont
  {{Salpeter}}(1977{\natexlab{a}})}]{1977ApJS...35..221S}%
  \BibitemOpen
  \bibfield  {author} {\bibinfo {author} {\bibfnamefont {D.~J.}\ \bibnamefont
  {{Stevenson}}}\ and\ \bibinfo {author} {\bibfnamefont {E.~E.}\ \bibnamefont
  {{Salpeter}}},\ }\bibfield  {title} {\bibinfo {title} {{The phase diagram and
  transport properties for hydrogen-helium fluid planets.}},\ }\href
  {https://doi.org/10.1086/190478} {\bibfield  {journal} {\bibinfo  {journal}
  {Astrophys. J. Suppl. Ser.}\ }\textbf {\bibinfo {volume} {35}},\ \bibinfo
  {pages} {221} (\bibinfo {year} {1977}{\natexlab{a}})}\BibitemShut {NoStop}%
\bibitem [{\citenamefont {{Stevenson}}\ and\ \citenamefont
  {{Salpeter}}(1977{\natexlab{b}})}]{1977ApJS...35..239S}%
  \BibitemOpen
  \bibfield  {author} {\bibinfo {author} {\bibfnamefont {D.~J.}\ \bibnamefont
  {{Stevenson}}}\ and\ \bibinfo {author} {\bibfnamefont {E.~E.}\ \bibnamefont
  {{Salpeter}}},\ }\bibfield  {title} {\bibinfo {title} {{The dynamics and
  helium distribution in hydrogen-helium fluid planets.}},\ }\href
  {https://doi.org/10.1086/190479} {\bibfield  {journal} {\bibinfo  {journal}
  {Astrophys. J. Suppl. Ser.}\ }\textbf {\bibinfo {volume} {35}},\ \bibinfo
  {pages} {239} (\bibinfo {year} {1977}{\natexlab{b}})}\BibitemShut {NoStop}%
\bibitem [{\citenamefont {{French}}\ \emph {et~al.}(2012)\citenamefont
  {{French}}, \citenamefont {{Becker}}, \citenamefont {{Lorenzen}},
  \citenamefont {{Nettelmann}}, \citenamefont {{Bethkenhagen}}, \citenamefont
  {{Wicht}},\ and\ \citenamefont {{Redmer}}}]{2012ApJS..202....5F}%
  \BibitemOpen
  \bibfield  {author} {\bibinfo {author} {\bibfnamefont {M.}~\bibnamefont
  {{French}}}, \bibinfo {author} {\bibfnamefont {A.}~\bibnamefont {{Becker}}},
  \bibinfo {author} {\bibfnamefont {W.}~\bibnamefont {{Lorenzen}}}, \bibinfo
  {author} {\bibfnamefont {N.}~\bibnamefont {{Nettelmann}}}, \bibinfo {author}
  {\bibfnamefont {M.}~\bibnamefont {{Bethkenhagen}}}, \bibinfo {author}
  {\bibfnamefont {J.}~\bibnamefont {{Wicht}}},\ and\ \bibinfo {author}
  {\bibfnamefont {R.}~\bibnamefont {{Redmer}}},\ }\bibfield  {title} {\bibinfo
  {title} {{Ab Initio Simulations for Material Properties along the Jupiter
  Adiabat}},\ }\href {https://doi.org/10.1088/0067-0049/202/1/5} {\bibfield
  {journal} {\bibinfo  {journal} {Astrophys. J. Suppl. Ser.}\ }\textbf
  {\bibinfo {volume} {202}},\ \bibinfo {eid} {5} (\bibinfo {year}
  {2012})}\BibitemShut {NoStop}%
\bibitem [{\citenamefont {{Rosenblum}}\ \emph {et~al.}(2011)\citenamefont
  {{Rosenblum}}, \citenamefont {{Garaud}}, \citenamefont {{Traxler}},\ and\
  \citenamefont {{Stellmach}}}]{2011ApJ...731...66R}%
  \BibitemOpen
  \bibfield  {author} {\bibinfo {author} {\bibfnamefont {E.}~\bibnamefont
  {{Rosenblum}}}, \bibinfo {author} {\bibfnamefont {P.}~\bibnamefont
  {{Garaud}}}, \bibinfo {author} {\bibfnamefont {A.}~\bibnamefont
  {{Traxler}}},\ and\ \bibinfo {author} {\bibfnamefont {S.}~\bibnamefont
  {{Stellmach}}},\ }\bibfield  {title} {\bibinfo {title} {{Turbulent Mixing and
  Layer Formation in Double-diffusive Convection: Three-dimensional Numerical
  Simulations and Theory}},\ }\href
  {https://doi.org/10.1088/0004-637X/731/1/66} {\bibfield  {journal} {\bibinfo
  {journal} {Astrophys. J.}\ }\textbf {\bibinfo {volume} {731}},\ \bibinfo
  {eid} {66} (\bibinfo {year} {2011})}\BibitemShut {NoStop}%
\bibitem [{\citenamefont {{Mirouh}}\ \emph {et~al.}(2012)\citenamefont
  {{Mirouh}}, \citenamefont {{Garaud}}, \citenamefont {{Stellmach}},
  \citenamefont {{Traxler}},\ and\ \citenamefont
  {{Wood}}}]{2012ApJ...750...61M}%
  \BibitemOpen
  \bibfield  {author} {\bibinfo {author} {\bibfnamefont {G.~M.}\ \bibnamefont
  {{Mirouh}}}, \bibinfo {author} {\bibfnamefont {P.}~\bibnamefont {{Garaud}}},
  \bibinfo {author} {\bibfnamefont {S.}~\bibnamefont {{Stellmach}}}, \bibinfo
  {author} {\bibfnamefont {A.~L.}\ \bibnamefont {{Traxler}}},\ and\ \bibinfo
  {author} {\bibfnamefont {T.~S.}\ \bibnamefont {{Wood}}},\ }\bibfield  {title}
  {\bibinfo {title} {{A New Model for Mixing by Double-diffusive Convection
  (Semi-convection). I. The Conditions for Layer Formation}},\ }\href
  {https://doi.org/10.1088/0004-637X/750/1/61} {\bibfield  {journal} {\bibinfo
  {journal} {Astrophys. J.}\ }\textbf {\bibinfo {volume} {750}},\ \bibinfo
  {eid} {61} (\bibinfo {year} {2012})}\BibitemShut {NoStop}%
\bibitem [{\citenamefont {{Wood}}\ \emph {et~al.}(2013)\citenamefont {{Wood}},
  \citenamefont {{Garaud}},\ and\ \citenamefont
  {{Stellmach}}}]{2013ApJ...768..157W}%
  \BibitemOpen
  \bibfield  {author} {\bibinfo {author} {\bibfnamefont {T.~S.}\ \bibnamefont
  {{Wood}}}, \bibinfo {author} {\bibfnamefont {P.}~\bibnamefont {{Garaud}}},\
  and\ \bibinfo {author} {\bibfnamefont {S.}~\bibnamefont {{Stellmach}}},\
  }\bibfield  {title} {\bibinfo {title} {{A New Model for Mixing by
  Double-diffusive Convection (Semi-convection). II. The Transport of Heat and
  Composition through Layers}},\ }\href
  {https://doi.org/10.1088/0004-637X/768/2/157} {\bibfield  {journal} {\bibinfo
   {journal} {Astrophys. J.}\ }\textbf {\bibinfo {volume} {768}},\ \bibinfo
  {eid} {157} (\bibinfo {year} {2013})}\BibitemShut {NoStop}%
\bibitem [{\citenamefont {{Moll}}\ \emph {et~al.}(2016)\citenamefont {{Moll}},
  \citenamefont {{Garaud}},\ and\ \citenamefont
  {{Stellmach}}}]{2016ApJ...823...33M}%
  \BibitemOpen
  \bibfield  {author} {\bibinfo {author} {\bibfnamefont {R.}~\bibnamefont
  {{Moll}}}, \bibinfo {author} {\bibfnamefont {P.}~\bibnamefont {{Garaud}}},\
  and\ \bibinfo {author} {\bibfnamefont {S.}~\bibnamefont {{Stellmach}}},\
  }\bibfield  {title} {\bibinfo {title} {{A New Model for Mixing by
  Double-diffusive Convection (Semi-convection). III. Thermal and Compositional
  Transport through Non-layered ODDC}},\ }\href
  {https://doi.org/10.3847/0004-637X/823/1/33} {\bibfield  {journal} {\bibinfo
  {journal} {Astrophys. J.}\ }\textbf {\bibinfo {volume} {823}},\ \bibinfo
  {eid} {33} (\bibinfo {year} {2016})}\BibitemShut {NoStop}%
\bibitem [{\citenamefont {{Herwig}}\ \emph {et~al.}(1997)\citenamefont
  {{Herwig}}, \citenamefont {{Bloecker}}, \citenamefont {{Schoenberner}},\ and\
  \citenamefont {{El Eid}}}]{1997A&A...324L..81H}%
  \BibitemOpen
  \bibfield  {author} {\bibinfo {author} {\bibfnamefont {F.}~\bibnamefont
  {{Herwig}}}, \bibinfo {author} {\bibfnamefont {T.}~\bibnamefont
  {{Bloecker}}}, \bibinfo {author} {\bibfnamefont {D.}~\bibnamefont
  {{Schoenberner}}},\ and\ \bibinfo {author} {\bibfnamefont {M.}~\bibnamefont
  {{El Eid}}},\ }\bibfield  {title} {\bibinfo {title} {{Stellar evolution of
  low and intermediate-mass stars. IV. Hydrodynamically-based overshoot and
  nucleosynthesis in AGB stars.}},\ }\href@noop {} {\bibfield  {journal}
  {\bibinfo  {journal} {Astron. Astrophys.}\ }\textbf {\bibinfo {volume}
  {324}},\ \bibinfo {pages} {L81} (\bibinfo {year} {1997})}\BibitemShut
  {NoStop}%
\bibitem [{\citenamefont {{Anders}}\ \emph {et~al.}(2022)\citenamefont
  {{Anders}}, \citenamefont {{Jermyn}}, \citenamefont {{Lecoanet}},\ and\
  \citenamefont {{Brown}}}]{2022ApJ...926..169A}%
  \BibitemOpen
  \bibfield  {author} {\bibinfo {author} {\bibfnamefont {E.~H.}\ \bibnamefont
  {{Anders}}}, \bibinfo {author} {\bibfnamefont {A.~S.}\ \bibnamefont
  {{Jermyn}}}, \bibinfo {author} {\bibfnamefont {D.}~\bibnamefont
  {{Lecoanet}}},\ and\ \bibinfo {author} {\bibfnamefont {B.~P.}\ \bibnamefont
  {{Brown}}},\ }\bibfield  {title} {\bibinfo {title} {{Stellar Convective
  Penetration: Parameterized Theory and Dynamical Simulations}},\ }\href
  {https://doi.org/10.3847/1538-4357/ac408d} {\bibfield  {journal} {\bibinfo
  {journal} {Astrophys. J.}\ }\textbf {\bibinfo {volume} {926}},\ \bibinfo
  {eid} {169} (\bibinfo {year} {2022})}\BibitemShut {NoStop}%
\bibitem [{\citenamefont {{Turner}}\ and\ \citenamefont
  {{Stommel}}(1964)}]{1964PNAS...52...49T}%
  \BibitemOpen
  \bibfield  {author} {\bibinfo {author} {\bibfnamefont {J.~S.}\ \bibnamefont
  {{Turner}}}\ and\ \bibinfo {author} {\bibfnamefont {H.}~\bibnamefont
  {{Stommel}}},\ }\bibfield  {title} {\bibinfo {title} {{A New Case of
  Convection in the Presence of Combined Vertical Salinity and Temperature
  Gradients}},\ }\href {https://doi.org/10.1073/pnas.52.1.49} {\bibfield
  {journal} {\bibinfo  {journal} {Proc. Natl. Acad. Sci. USA}\ }\textbf
  {\bibinfo {volume} {52}},\ \bibinfo {pages} {49} (\bibinfo {year}
  {1964})}\BibitemShut {NoStop}%
\bibitem [{\citenamefont {Turner}(1968)}]{turner_1968}%
  \BibitemOpen
  \bibfield  {author} {\bibinfo {author} {\bibfnamefont {J.~S.}\ \bibnamefont
  {Turner}},\ }\bibfield  {title} {\bibinfo {title} {The behaviour of a stable
  salinity gradient heated from below},\ }\href
  {https://doi.org/10.1017/S0022112068002442} {\bibfield  {journal} {\bibinfo
  {journal} {J. Fluid Mech}\ }\textbf {\bibinfo {volume} {33}},\ \bibinfo
  {pages} {183–200} (\bibinfo {year} {1968})}\BibitemShut {NoStop}%
\bibitem [{\citenamefont {Fernando}(1987)}]{fernando_1987}%
  \BibitemOpen
  \bibfield  {author} {\bibinfo {author} {\bibfnamefont {H.~J.~S.}\
  \bibnamefont {Fernando}},\ }\bibfield  {title} {\bibinfo {title} {The
  formation of a layered structure when a stable salinity gradient is heated
  from below},\ }\href {https://doi.org/10.1017/S0022112087002441} {\bibfield
  {journal} {\bibinfo  {journal} {J. Fluid Mech}\ }\textbf {\bibinfo {volume}
  {182}},\ \bibinfo {pages} {525–541} (\bibinfo {year} {1987})}\BibitemShut
  {NoStop}%
\bibitem [{\citenamefont {{Fuentes}}\ and\ \citenamefont
  {{Cumming}}(2020)}]{2020PhRvF...5l4501F}%
  \BibitemOpen
  \bibfield  {author} {\bibinfo {author} {\bibfnamefont {J.~R.}\ \bibnamefont
  {{Fuentes}}}\ and\ \bibinfo {author} {\bibfnamefont {A.}~\bibnamefont
  {{Cumming}}},\ }\bibfield  {title} {\bibinfo {title} {{Penetration of a
  cooling convective layer into a stably-stratified composition gradient:
  Entrainment at low Prandtl number}},\ }\href
  {https://doi.org/10.1103/PhysRevFluids.5.124501} {\bibfield  {journal}
  {\bibinfo  {journal} {Phys. Rev. Fluids}\ }\textbf {\bibinfo {volume} {5}},\
  \bibinfo {eid} {124501} (\bibinfo {year} {2020})}\BibitemShut {NoStop}%
\bibitem [{\citenamefont {{Spiegel}}\ and\ \citenamefont
  {{Veronis}}(1960)}]{1960ApJ...131..442S}%
  \BibitemOpen
  \bibfield  {author} {\bibinfo {author} {\bibfnamefont {E.~A.}\ \bibnamefont
  {{Spiegel}}}\ and\ \bibinfo {author} {\bibfnamefont {G.}~\bibnamefont
  {{Veronis}}},\ }\bibfield  {title} {\bibinfo {title} {{On the Boussinesq
  Approximation for a Compressible Fluid.}},\ }\href
  {https://doi.org/10.1086/146849} {\bibfield  {journal} {\bibinfo  {journal}
  {Astrophys. J.}\ }\textbf {\bibinfo {volume} {131}},\ \bibinfo {pages} {442}
  (\bibinfo {year} {1960})}\BibitemShut {NoStop}%
\bibitem [{\citenamefont {{Fitzgerald}}\ and\ \citenamefont
  {{Farrell}}(2014)}]{2014PhFl...26e4104F}%
  \BibitemOpen
  \bibfield  {author} {\bibinfo {author} {\bibfnamefont {J.~G.}\ \bibnamefont
  {{Fitzgerald}}}\ and\ \bibinfo {author} {\bibfnamefont {B.~F.}\ \bibnamefont
  {{Farrell}}},\ }\bibfield  {title} {\bibinfo {title} {{Mechanisms of mean
  flow formation and suppression in two-dimensional Rayleigh-B{\'e}nard
  convection}},\ }\href {https://doi.org/10.1063/1.4875814} {\bibfield
  {journal} {\bibinfo  {journal} {Phys. Fluids}\ }\textbf {\bibinfo {volume}
  {26}},\ \bibinfo {eid} {054104} (\bibinfo {year} {2014})}\BibitemShut
  {NoStop}%
\bibitem [{\citenamefont {{Wang}}\ \emph {et~al.}(2020)\citenamefont {{Wang}},
  \citenamefont {{Chong}}, \citenamefont {{Stevens}}, \citenamefont
  {{Verzicco}},\ and\ \citenamefont {{Lohse}}}]{2020JFM...905A..21W}%
  \BibitemOpen
  \bibfield  {author} {\bibinfo {author} {\bibfnamefont {Q.}~\bibnamefont
  {{Wang}}}, \bibinfo {author} {\bibfnamefont {K.~L.}\ \bibnamefont {{Chong}}},
  \bibinfo {author} {\bibfnamefont {R.~J.~A.~M.}\ \bibnamefont {{Stevens}}},
  \bibinfo {author} {\bibfnamefont {R.}~\bibnamefont {{Verzicco}}},\ and\
  \bibinfo {author} {\bibfnamefont {D.}~\bibnamefont {{Lohse}}},\ }\bibfield
  {title} {\bibinfo {title} {{From zonal flow to convection rolls in
  {R}ayleigh-{B}{\'e}nard convection with free-slip plates}},\ }\href
  {https://doi.org/10.1017/jfm.2020.793} {\bibfield  {journal} {\bibinfo
  {journal} {J. Fluid Mech.}\ }\textbf {\bibinfo {volume} {905}},\ \bibinfo
  {eid} {A21} (\bibinfo {year} {2020})}\BibitemShut {NoStop}%
\bibitem [{\citenamefont {{Fuentes}}\ and\ \citenamefont
  {{Cumming}}(2021)}]{2021PhRvF...6g4502F}%
  \BibitemOpen
  \bibfield  {author} {\bibinfo {author} {\bibfnamefont {J.~R.}\ \bibnamefont
  {{Fuentes}}}\ and\ \bibinfo {author} {\bibfnamefont {A.}~\bibnamefont
  {{Cumming}}},\ }\bibfield  {title} {\bibinfo {title} {{Shear flows and their
  suppression at large aspect ratio: Two-dimensional simulations of a growing
  convection zone}},\ }\href {https://doi.org/10.1103/PhysRevFluids.6.074502}
  {\bibfield  {journal} {\bibinfo  {journal} {Physical Review Fluids}\ }\textbf
  {\bibinfo {volume} {6}},\ \bibinfo {eid} {074502} (\bibinfo {year}
  {2021})}\BibitemShut {NoStop}%
\bibitem [{\citenamefont {{Burns}}\ \emph {et~al.}(2020)\citenamefont
  {{Burns}}, \citenamefont {{Vasil}}, \citenamefont {{Oishi}}, \citenamefont
  {{Lecoanet}},\ and\ \citenamefont {{Brown}}}]{2020PhRvR...2b3068B}%
  \BibitemOpen
  \bibfield  {author} {\bibinfo {author} {\bibfnamefont {K.~J.}\ \bibnamefont
  {{Burns}}}, \bibinfo {author} {\bibfnamefont {G.~M.}\ \bibnamefont
  {{Vasil}}}, \bibinfo {author} {\bibfnamefont {J.~S.}\ \bibnamefont
  {{Oishi}}}, \bibinfo {author} {\bibfnamefont {D.}~\bibnamefont
  {{Lecoanet}}},\ and\ \bibinfo {author} {\bibfnamefont {B.~P.}\ \bibnamefont
  {{Brown}}},\ }\bibfield  {title} {\bibinfo {title} {{Dedalus: A flexible
  framework for numerical simulations with spectral methods}},\ }\href
  {https://doi.org/10.1103/PhysRevResearch.2.023068} {\bibfield  {journal}
  {\bibinfo  {journal} {Phys. Rev. Research}\ }\textbf {\bibinfo {volume}
  {2}},\ \bibinfo {eid} {023068} (\bibinfo {year} {2020})}\BibitemShut
  {NoStop}%
\bibitem [{\citenamefont {Ascher}\ \emph {et~al.}(1997)\citenamefont {Ascher},
  \citenamefont {Ruuth},\ and\ \citenamefont {Spiteri}}]{ASCHER1997151}%
  \BibitemOpen
  \bibfield  {author} {\bibinfo {author} {\bibfnamefont {U.~M.}\ \bibnamefont
  {Ascher}}, \bibinfo {author} {\bibfnamefont {S.~J.}\ \bibnamefont {Ruuth}},\
  and\ \bibinfo {author} {\bibfnamefont {R.~J.}\ \bibnamefont {Spiteri}},\
  }\bibfield  {title} {\bibinfo {title} {Implicit-explicit runge-kutta methods
  for time-dependent partial differential equations},\ }\href
  {https://doi.org/https://doi.org/10.1016/S0168-9274(97)00056-1} {\bibfield
  {journal} {\bibinfo  {journal} {Appl. Numer. Math.}\ }\textbf {\bibinfo
  {volume} {25}},\ \bibinfo {pages} {151} (\bibinfo {year} {1997})},\ \bibinfo
  {note} {special Issue on Time Integration}\BibitemShut {NoStop}%
\bibitem [{\citenamefont {{Jeroen Molemaker}}\ and\ \citenamefont
  {{Dijkstra}}(1997)}]{1997JFM...331..199J}%
  \BibitemOpen
  \bibfield  {author} {\bibinfo {author} {\bibfnamefont {M.}~\bibnamefont
  {{Jeroen Molemaker}}}\ and\ \bibinfo {author} {\bibfnamefont {H.~A.}\
  \bibnamefont {{Dijkstra}}},\ }\bibfield  {title} {\bibinfo {title} {{The
  formation and evolution of a diffusive interface}},\ }\href@noop {}
  {\bibfield  {journal} {\bibinfo  {journal} {J. Fluid Mech.}\ }\textbf
  {\bibinfo {volume} {331}},\ \bibinfo {pages} {199} (\bibinfo {year}
  {1997})}\BibitemShut {NoStop}%
\bibitem [{\citenamefont {{Zaussinger}}\ and\ \citenamefont
  {{Kupka}}(2019)}]{2019ThCFD..33..383Z}%
  \BibitemOpen
  \bibfield  {author} {\bibinfo {author} {\bibfnamefont {F.}~\bibnamefont
  {{Zaussinger}}}\ and\ \bibinfo {author} {\bibfnamefont {F.}~\bibnamefont
  {{Kupka}}},\ }\bibfield  {title} {\bibinfo {title} {{Layer formation in
  double-diffusive convection over resting and moving heated plates}},\ }\href
  {https://doi.org/10.1007/s00162-019-00499-7} {\bibfield  {journal} {\bibinfo
  {journal} {Theor. Comput. Fluid Dyn.}\ }\textbf {\bibinfo {volume} {33}},\
  \bibinfo {pages} {383} (\bibinfo {year} {2019})}\BibitemShut {NoStop}%
\bibitem [{\citenamefont {{Schmalzl}}\ \emph {et~al.}(2004)\citenamefont
  {{Schmalzl}}, \citenamefont {{Breuer}},\ and\ \citenamefont
  {{Hansen}}}]{2004EL.....67..390S}%
  \BibitemOpen
  \bibfield  {author} {\bibinfo {author} {\bibfnamefont {J.}~\bibnamefont
  {{Schmalzl}}}, \bibinfo {author} {\bibfnamefont {M.}~\bibnamefont
  {{Breuer}}},\ and\ \bibinfo {author} {\bibfnamefont {U.}~\bibnamefont
  {{Hansen}}},\ }\bibfield  {title} {\bibinfo {title} {{On the validity of
  two-dimensional numerical approaches to time-dependent thermal convection}},\
  }\href {https://doi.org/10.1209/epl/i2003-10298-4} {\bibfield  {journal}
  {\bibinfo  {journal} {Europhys. Lett.}\ }\textbf {\bibinfo {volume} {67}},\
  \bibinfo {pages} {390} (\bibinfo {year} {2004})}\BibitemShut {NoStop}%
\end{thebibliography}%


%
\end{document}